\documentclass[a4paper,11pt]{article}
\pdfoutput=1
\usepackage{jheppub} 
\usepackage{bm} 
\usepackage{bbm}
\usepackage{mathtools} 
\usepackage[T1]{fontenc} 
\usepackage[normalem]{ulem} 

\usepackage[pdftex]{color}
\usepackage{xcolor}

\allowdisplaybreaks 

\title{Inclusive Production of Heavy Quarkonia in pNRQCD} 
\preprint{TUM-EFT 139/20} 

\author[a,b]{Nora~Brambilla,} 
\author[a,c]{Hee~Sok~Chung,} 
\author[a]{and Antonio~Vairo} 

\affiliation[a]{Physik-Department, Technische Universit\"at M\"unchen, James-Franck-Str. 1, 85748 Garching, Germany} 
\affiliation[b]{Institute for Advanced Study, Technische Universit\"at M\"unchen, Lichtenbergstrasse 2~a, 85748 Garching, Germany} 
\affiliation[c]{Excellence Cluster ORIGINS, Boltzmannstrasse 2, 85748 Garching, Germany} 

\emailAdd{nora.brambilla@tum.de} 
\emailAdd{heesok.chung@tum.de} 
\emailAdd{antonio.vairo@tum.de} 

\abstract{We develop a formalism for computing inclusive production cross sections of
  heavy quarkonia based on the nonrelativistic QCD and the potential nonrelativistic QCD effective field theories.
  Our formalism applies to strongly coupled quarkonia, which include excited charmonium and bottomonium states.
  Analogously to heavy quarkonium decay processes, we express nonrelativistic QCD long-distance matrix elements in terms of quarkonium wavefunctions
  at the origin and universal gluonic correlators.
  Our expressions for the long-distance matrix elements are valid up to corrections of order $1/N_c^2$.
  These expressions enhance the predictive power of the nonrelativistic effective field theory approach to inclusive production processes
  by reducing the number of nonperturbative unknowns, and make possible first-principle determinations of long-distance matrix elements once the gluonic correlators are known.
  Based on this formalism, we compute the production cross sections of $P$-wave charmonia and bottomonia at the LHC, and find good agreement with measurements.  
}

\begin{document} 
\maketitle 
\flushbottom

\section{Introduction} 
Understanding the mechanism of inclusive heavy quarkonium production is one of the most challenging problems
in QCD~\cite{Brambilla:2004wf, Brambilla:2010cs, Bodwin:2013nua, Brambilla:2014jmp}.
Heavy quarkonium production processes provide probes of the interplay between perturbative and nonperturbative aspects of QCD,
and are considered key in understanding the hot and dense quark-gluon plasma.  
For decades, much theoretical effort has been made using the nonrelativistic QCD (NRQCD) factorization formalism~\cite{Bodwin:1994jh}.
In this formalism, inclusive production rates of a heavy quarkonium are factorized into products of short-distance coefficients,
which encode the perturbative physics at the scale of the heavy quark mass and above,
and long-distance matrix elements (LDMEs) that depend on the nonperturbative nature of the quarkonium state.  
While much progress has been made in computing the short-distance coefficients in perturbative QCD, 
it remains unknown how to compute from first principles a wide class of LDMEs.
Therefore, in most phenomenological studies, the long-distance matrix elements have been obtained from fits to cross section data.  
This approach has resulted in different sets of LDME determinations depending on the choice of data that are used in the fit,
which can disagree with one another~\cite{Chung:2018lyq}.  
These inconsistent sets of LDMEs lead to contradicting predictions,
and none of the determinations is able to give a comprehensive description of the most important observables related
to heavy quarkonium production at a satisfactory level~\cite{Chung:2018lyq,Lansberg:2019adr}.  
It is therefore desirable and to some extent even necessary to have first-principle based constraints, or even computations, of the NRQCD LDMEs,
in order to enhance significantly our understanding of the heavy quarkonium production mechanism.  

Nonrelativistic effective field theories such as NRQCD exploit the hierarchy of energy scales that appear in processes involving heavy quarkonia,
which are the heavy quark mass $m$, the typical relative momentum $mv$ of the quark and antiquark, and the typical binding energy $mv^2$~\cite{Brambilla:2004jw}.
Here, $v \ll 1$ is the relative velocity of the heavy quark and the antiquark inside a heavy quarkonium.\footnote{
  Reference values for $v^2$ are 0.3 in the charmonium case and 0.1 in the bottomonium one.
  We will use these values in phenomenological applications when estimating the uncertainties due to higher order terms in the $v$ expansion. 
}  
NRQCD is obtained by integrating out modes associated with energy scales of order $m$ and higher~\cite{Caswell:1985ui,Bodwin:1994jh}.
These are encoded in the short-distance coefficients,  while all contributions from the scales $mv$ and $mv^2$ are contained in the NRQCD LDMEs.
Because the heavy quark mass is much larger than $\Lambda_{\rm QCD}$, the short-distance coefficients can be computed in perturbative QCD, and are given by series in $\alpha_{\text{s}}$.
The LDMEs that appear in the NRQCD factorization formulas for inclusive production cross sections of heavy quarkonia describe the 
evolution of a heavy quark and antiquark pair into a heavy quarkonium state.  
This should be contrasted with the LDMEs for decay rates of heavy quarkonia that correspond to probabilities of finding a heavy quark and antiquark pair inside a heavy quarkonium.
In case of LDMEs that involve heavy quark and antiquark pairs in color-singlet states,
the so-called vacuum-saturation approximation can be used to relate the production and decay LDMEs.  
Hence, the color-singlet production LDMEs have been obtained from lattice QCD, potential-model calculations, or from measured decay rates.  
On the other hand, it is not known how to relate the production and decay LDMEs that involve heavy quark and antiquark pairs in color-octet states,
and the color-octet production LDMEs have not been computed from first principles.  

The effective field theory potential NRQCD (pNRQCD)~\cite{Pineda:1997bj,Brambilla:1999xf,Brambilla:2000gk,Pineda:2000sz,Brambilla:2004jw}
is obtained by further integrating out modes associated with energy scales larger than $mv^2$ appearing in NRQCD.
In the regime $mv^2 \ll \Lambda_{\rm QCD}$, which is satisfied by non-Coulombic, strongly coupled quarkonia,
NRQCD decay LDMEs can be factorized into products of quarkonium wavefunctions at the origin and universal gluonic correlators.
The quarkonium wavefunctions are determined by solving the Schr\"odinger equation, which is the equation of motion of pNRQCD at leading order in $v$.
The gluonic correlators are universal quantities that can be computed in lattice QCD.
Strongly coupled potential NRQCD has been successfully applied to heavy quarkonium decay and exclusive electromagnetic production
processes~\cite{Brambilla:2001xy,Brambilla:2002nu,Brambilla:2003mu,Brambilla:2020xod}.
It has been anticipated that a similar formalism could be used to describe inclusive production processes of heavy quarkonia, by computing the production LDMEs in pNRQCD.
Such a calculation would allow model-independent, first-principle based predictions of heavy quarkonium production cross sections.  

Calculations of production LDMEs are also important in understanding the NRQCD factorization formalism for inclusive production of heavy quarkonia.  
The validity of the factorization formalism depends on the infrared finiteness of the short-distance coefficients to all orders in the expansion in powers of $\alpha_{\text{s}}$.
An essential ingredient in proving such factorization is the determination of the infrared behavior of the production LDMEs.
So far, the infrared properties of the production LDMEs have only been studied to two-loop accuracy~\cite{Nayak:2005rw,Nayak:2005rt,Nayak:2006fm,
Bodwin:2019bpf, Zhang:2020atv}.  
It is possible that the pNRQCD expressions of the production LDMEs will simplify the investigation of their infrared properties and the verification of the NRQCD factorization formalism.  

We have presented a first calculation of production LDMEs in strongly coupled pNRQCD in ref.~\cite{Brambilla:2020ojz} for production of $P$-wave heavy quarkonia.
In this paper, we describe in detail the formalism and clarify some technical details that may be useful in future studies.
We work in the strong coupling regime $mv^2 \ll \Lambda_{\rm QCD}$, and with heavy quarkonium states that are below the open flavor threshold.  
The calculation of the production LDMEs in this paper is valid up to corrections of relative order $1/N_c^2$ and $v^2$.\footnote{
In the special kinematical situation  $mv^2 \ll \Lambda_{\rm QCD} \ll mv$, the ratio $mv^2/\Lambda_{\rm QCD}$ is larger than $v$ 
and the induced corrections to the LDMEs may turn out to be parametrically larger than $v^2$, which is their natural size when $\Lambda_{\rm QCD} \sim mv$.
In order not to complicate unnecessarily the error estimate, in the rest of the paper we will count the neglected corrections to the LDMEs in the $v$ expansion as $v^2$ for all the kinematical 
situations that fulfill $mv^2 \ll \Lambda_{\rm QCD}$.
}
Furthermore, we will present an extended set of phenomenological results.

The paper is organized as follows.
In section~\ref{sec:ldmes}, we set up the general formalism for the computation of the  production LDMEs in pNRQCD. 
In section~\ref{sec:pwave-th}, we express in pNRQCD the production LDMEs for $P$-wave heavy quarkonium states.
Phenomenological applications that include the computation of the cross sections for $\chi_{cJ}(1P)$ and $\chi_{bJ}(nP)$ states at the LHC can be found in section~\ref{sec:pwave-phen}.
We conclude in section~\ref{sec:conclusion}.

\section{LDMEs in pNRQCD} 
\label{sec:ldmes} 
In the NRQCD factorization formalism, the inclusive production cross section of a quarkonium ${\cal Q}$ is given by~\cite{Bodwin:1994jh} 
\begin{equation} 
\label{eq:NRQCDfac} 
\sigma_{{\cal Q}+X} =  \sum_N \sigma_{Q \bar Q(N)} \langle \Omega | {\cal O}^{\cal Q} (N) | \Omega \rangle.  
\end{equation} 
Here, $\sigma_{Q \bar Q(N)}$ are short-distance coefficients that correspond to the production cross section
of a heavy quark-antiquark pair ($Q \bar Q$) in a spin and color state $N$, and $|\Omega \rangle$ is the QCD vacuum state.
The NRQCD long-distance matrix element $\langle \Omega | {\cal O}^{\cal Q} (N) | \Omega \rangle$ describes the evolution of the $Q \bar Q$ in a state $N$ into the quarkonium ${\cal Q}$.  
The matrix elements have known scalings in $v$, which allows us to organize
the sum over $N$ in eq.~\eqref{eq:NRQCDfac} in powers of $v$.  
In general, the $Q \bar Q$ can be in a color-singlet or in a color-octet state.  

For a color-singlet state, the operators ${\cal O}^{\cal Q} (N)$ have the form 
\begin{equation} 
\label{eq:cs_ldme} 
{\cal O}^{\cal Q} (N_{\textrm{color singlet}}) = \chi^\dag {\cal K}_N \psi {\cal P}_{{\cal Q}(\bm{P}=\bm{0})} \psi^\dag {\cal K}'_N \chi, 
\end{equation} 
where $\psi$ and $\chi$ are Pauli spinor fields that annihilate and create a heavy quark and antiquark at spacetime position $0$, respectively.
The operator ${\cal P}_{{\cal Q} (\bm{P})}$ projects onto a state consisting of a heavy quarkonium $\cal Q$ with momentum $\bm{P}$;
it may be written as ${\cal P}_{{\cal Q}(\bm{P})} = a_{{\cal Q}(\bm{P})}^\dag a_{{\cal Q}(\bm{P})}$,
where $a_{{\cal Q}(\bm{P})}^\dag$ is an operator that creates a quarkonium ${\cal Q}$ with momentum $\bm{P}$.  
The quantities ${\cal K}_N$ and ${\cal K}'_N$ are polynomials of Pauli matrices and covariant derivatives, and are proportional to the color identity matrix $\mathbbm{1}_c$,
so that the operators $\chi^\dag {\cal K}_N \psi$ and $\psi^\dag {\cal K}'_N \chi$ are color singlets.  

For a color-octet state, the operators ${\cal O}^{\cal Q} (N)$ take the form~\cite{Nayak:2005rw,Nayak:2005rt,Nayak:2006fm} 
\begin{equation} 
\label{eq:co_ldme} 
{\cal O}^{\cal Q} (N_{\textrm{color octet}}) = \chi^\dag {\cal K}_N T^a \psi \Phi_\ell^{\dag ab} (0) {\cal P}_{{\cal Q}(\bm{P}=\bm{0})} \Phi_\ell^{bc} (0) \psi^\dag {\cal K}'_N T^c \chi, 
\end{equation} 
where $T^a$ is a color matrix in the fundamental representation.
The operator $\Phi_\ell(x)$ is a Wilson line along the direction $\ell$ in the adjoint representation defined by 
\begin{align} 
\Phi_\ell(x) = {\cal P} \exp \left[ -i g \int_0^\infty d \lambda \, \ell \cdot A^{\rm adj} (x+\ell \lambda) \right],
\end{align} 
where $A^{\rm adj}$ is the gluon field in the adjoint representation and ${\cal P}$ stands for the path ordering of the color matrices.
The Wilson lines $\Phi_\ell$ are necessary to ensure the gauge invariance of the vacuum expectation value of eq.~\eqref{eq:co_ldme}, because in the absence of $\Phi_\ell$,
the gauge transformations of $\chi^\dag {\cal K}_N T^a \psi$ and $\psi^\dag {\cal K}'_N T^c \chi$ do not commute with the operator ${\cal P}_{{\cal Q}(\bm{P}=\bm{0})}$.  
The direction $\ell$ is arbitrary.  
Hereafter we refer to the $\Phi_\ell$ in eq.~\eqref{eq:co_ldme} as gauge-completion Wilson lines.  

The NRQCD factorization conjecture is the statement that the  short-distance coefficients in eq.~\eqref{eq:NRQCDfac} can be computed in perturbative QCD,
and the NRQCD matrix elements are universal.  
The  short-distance coefficients are computed in perturbative QCD by replacing the heavy quarkonium state ${\cal Q}$ in eq.~\eqref{eq:NRQCDfac} with a perturbative $Q \bar Q$ state.  
For the  short-distance coefficients to be perturbatively calculable, the infrared (IR) divergences that appear in the perturbative QCD calculation of 
eq.~\eqref{eq:NRQCDfac} must be reproduced by the NRQCD matrix elements to all orders in $\alpha_{\text{s}}$, so that the short-distance coefficients  are IR finite.  
The outline for an all-orders proof of the IR finiteness of the short-distance coefficients has been given in refs.~\cite{Nayak:2005rt,Kang:2014tta}
for the case where the quarkonium ${\cal Q}$ is produced with a momentum that exceeds the mass of the heavy quarkonium $m_{\cal Q}$: 
for example, the proof applies to the case when the quarkonium is produced with a large transverse momentum $p_T \gg m_{\cal Q}$
and the cross section is expanded to next-to-leading power in $m_{\cal
Q}^2/p_T^2$.  
On the other hand, the universality of the NRQCD matrix elements requires the color-octet matrix elements to be independent of the direction of the gauge-completion Wilson lines.  
A proof of the universality based on traditional methods of perturbative factorization has only been investigated to next-to-next-to-leading order 
in $\alpha_{\text{s}}$ for the specific case where ${\cal K}_N = {\cal K}'_N =
1$~\cite{Nayak:2005rw, Nayak:2005rt,Nayak:2006fm, Bodwin:2019bpf, Zhang:2020atv}.

In this paper, we aim at expressing the color-singlet and color-octet long distance matrix elements in pNRQCD.
We work in the strong coupling regime, where $mv^2 \ll \Lambda_{\rm QCD}$.
This condition is fulfilled by non-Coulombic, strongly coupled quarkonia.
The computation is based on expanding the NRQCD Hamiltonian in inverse powers of the heavy quark mass~$m$,
\begin{equation} 
\label{QMPT1} 
H_{\rm NRQCD} = H_{\rm NRQCD}^{(0)} + \frac{H_{\rm NRQCD}^{(1)}}{m} + \ldots , 
\end{equation} 
and on making use of quantum-mechanical perturbation theory to compute its eigenstates order by order in $1/m$~\cite{Brambilla:2000gk,Pineda:2000sz}.
Because the energy scales that appear in NRQCD are $mv$, $mv^2$, and $\Lambda_{\rm QCD}$, 
the expansion in powers of $1/m$ in the Hamiltonian leads to an expansion in powers of the dimensionless parameters 
$v$ and $\Lambda_{\rm QCD}/m$ in the observables.
Explicitly, the static NRQCD Hamiltonian, $H_{\rm NRQCD}^{(0)}$, and its $1/m$ correction $H_{\rm NRQCD}^{(1)}$ are given by 
\begin{align} 
H_{\rm NRQCD}^{(0)} =& 
\, \frac{1}{2} \int d^3x \, (\bm{E}^a \cdot \bm{E}^a + \bm{B}^a \cdot \bm{B}^a) - \sum_{k=1}^{n_f} \int d^3x \, \bar q_k i \bm{D} \cdot \bm{\gamma} q_k, 
\\ 
H_{\rm NRQCD}^{(1)} =& 
- \frac{1}{2} \int d^3x \, \psi^\dag \bm{D}^2 \psi - \frac{c_F}{2} \int d^3x \, \psi^\dag \bm{\sigma} \cdot g \bm{B} \psi 
\nonumber \\ 
& + \frac{1}{2} \int d^3x \, \chi^\dag \bm{D}^2 \chi+ \frac{c_F}{2} \int d^3x \, \chi^\dag \bm{\sigma} \cdot g \bm{B} \chi,
\end{align} 
where $E^{i\,a} T^a = E^i = G^{i0}$ and $B^{i \,a} T^a = B^i = - \epsilon_{ijk} G^{jk}/2$ are the chromoelectric and chromomagnetic fields, respectively, 
$G^{\mu \nu \, a} T^a = G^{\mu \nu}$ is the gluon field strength tensor, $q_k$ are $n_f$ massless quark fields, and $\bm{D} = \bm{\nabla}-ig \bm{A}$ is the gauge covariant derivative.
The matrices $\sigma^i$ are the Pauli matrices, and $c_F$ is a short-distance coefficient, which is known to three-loop accuracy~\cite{Grozin:2007fh}.  
The physical states $|\rm phys\rangle$ are constrained by the Gauss law,
which reads
\begin{equation}
\bm{D} \cdot \bm{\Pi}^a | {\rm phys} \rangle
= g \left( \psi^\dag T^a \psi + \chi^\dag T^a \chi + \sum_{k=1}^{n_f} \bar{q}_k
\gamma^0 T^a q_k \right) | {\rm phys} \rangle, 
\end{equation}
where $\bm{\Pi}^a$($\approx \bm{E}^a$) is the canonical momentum conjugated to $\bm{A}^a$.

The space of states of NRQCD in the $Q \bar Q$ sector is spanned by states $|\underline{\rm n}; \bm{x}_1, \bm{x}_2 \rangle$
containing a heavy quark located at $\bm{x}_1$ and a heavy antiquark located at $\bm{x}_2$:
\begin{equation} 
\label{nbarvsn}
|\underline{\rm n}; \bm{x}_1, \bm{x}_2 \rangle = \psi^\dag(\bm{x}_1) \chi(\bm{x}_2) | n; \bm{x}_1, \bm{x}_2 \rangle, 
\end{equation} 
where the states $|n; \bm{x}_1, \bm{x}_2 \rangle$ do not contain heavy particles.\footnote{\label{footindex}
  For further use we make explicit the color indices of the state:
  $$
  |\underline{\rm n}; \bm{x}_1, \bm{x}_2 \rangle = \psi_i^\dag(\bm{x}_1) \chi_j(\bm{x}_2) | n; \bm{x}_1, \bm{x}_2;i,j \rangle.
  $$
}
We may take both $|\underline{\rm n}; \bm{x}_1, \bm{x}_2 \rangle$ and $| n; \bm{x}_1, \bm{x}_2 \rangle$ to satisfy  orthonormality relations.
Furthermore, the states $|\underline{\rm n}; \bm{x}_1, \bm{x}_2 \rangle$ are such that the NRQCD Hamiltonian is diagonal in $n$ on them:
\begin{equation}
\langle\underline{\rm n}; \bm{x}_1, \bm{x}_2 |H_{\rm NRQCD} ,|\underline{\rm m}; \bm{x}'_1, \bm{x}'_2 \rangle =
\delta_{nm}\, E_n(\bm{x}_1, \bm{x}_2; \bm{\nabla}_1, \bm{\nabla}_2)\, \delta^{(3)}(\bm{x}_1 - \bm{x}'_1)\, \delta^{(3)}(\bm{x}_2 - \bm{x}'_2)\,,
\label{braHket}
\end{equation}  
where $\bm{\nabla}_i = \bm{\nabla}_{x_i}$.
Exploiting the expansion \eqref{QMPT1}, the states $|\underline{\rm n}; \bm{x}_1, \bm{x}_2 \rangle$ can be computed order by order in $1/m$: 
\begin{equation} 
\label{QMPT2} 
|\underline{\rm n}; \bm{x}_1, \bm{x}_2 \rangle = |\underline{\rm n}; \bm{x}_1, \bm{x}_2 \rangle^{(0)} + \frac{|\underline{\rm n}; \bm{x}_1, \bm{x}_2 \rangle^{(1)}}{m} + \ldots.
\end{equation} 
The states $|\underline{\rm n}; \bm{x}_1, \bm{x}_2\rangle^{(0)}$ are eigenstates of $H_{\rm NRQCD}^{(0)}$, whose eigenvalues are the static energies $E^{(0)}_n( \bm{x}_1, \bm{x}_2)$;
the coordinates $\bm{x}_1$ and $\bm{x}_2$ are conserved in the static limit.
The states $|\underline{\rm n}; \bm{x}_1, \bm{x}_2 \rangle^{(1)}/m$ are the order-$1/m$ corrections due to $H_{\rm NRQCD}^{(1)}/m$.
The explicit expression of $|\underline{\rm n}; \bm{x}_1, \bm{x}_2 \rangle^{(1)}$ in terms of $|\underline{\rm n}; \bm{x}_1, \bm{x}_2 \rangle^{(0)}$ and $E^{(0)}_n( \bm{x}_1, \bm{x}_2)$
can be found in refs.~\cite{Brambilla:2000gk,Pineda:2000sz}.  

The number $n$ labels the QCD static energies $E^{(0)}_n( \bm{x}_1, \bm{x}_2)$. 
Each static energy describes physically a distinct excitation of the static quark-antiquark pair due to the light degrees of freedom (gluons, light quarks).
Lattice calculations of the QCD static spectrum suggest that 
each excitation is separated from the other by an energy gap of order $mv$ or $\Lambda_{\rm QCD}$~\cite{Bali:2000vr,Juge:2002br,Capitani:2018rox}.
The $n=0$ state corresponds to the ground state.
For vanishing heavy quark-antiquark distance, $\bm{r} = \bm{x}_1 - \bm{x}_2$, the ground state reduces to the QCD vacuum:
\begin{equation} 
\label{0vsOmega}
\delta^{(3)}(\bm{r}) |0; \bm{x}_1, \bm{x}_2 \rangle^{(0)} = \delta^{(3)}(\bm{r}) \frac{\mathbbm{1}_c}{\sqrt{N_c}} |\Omega \rangle,
\end{equation} 
where $N_c$ is the number of colors.
This follows from the fact that the gluonic component of the Fock state describing a static heavy quark-antiquark pair in the ground state located at the origin
is proportional to the color identity, the state being a color singlet,  and is proportional to the QCD vacuum state, the state being the ground state.

We compute the matrix elements $\langle \Omega | {\cal O}^{\cal Q} (N) | \Omega \rangle$ by first evaluating the matrix elements of the operator ${\cal P}_{{\cal Q} (\bm{P}=\bm{0})}$ 
on the states $|\underline{\rm n}; \bm{x}_1, \bm{x}_2\rangle$ in pNRQCD, which we express in terms of quarkonium wavefunctions.  
Then, the remaining matrix elements of the heavy quark and gluon operators can be evaluated by using the methods first developed in refs.~\cite{Brambilla:2001xy,Brambilla:2002nu}
for computing the decay matrix elements in pNRQCD.

\subsection[Matching of ${\cal P}_{{\cal Q}({\bf\it P})}$]{\boldmath Matching of ${\cal P}_{{\cal Q}({\bf\it P})}$} 
In this section, we express the matrix elements of the operator ${\cal P}_{{\cal Q}(\bm{P})} = a_{{\cal Q}(\bm{P})}^\dag a_{{\cal Q}(\bm{P})}$
between states $|\underline{\rm n}; \bm{x}_1, \bm{x}_2\rangle$ in terms of quarkonium wavefunctions.
Our computation is based on some general properties of the operator.  

In NRQCD, processes that involve momentum transfers at the scale $m$ are integrated out.
Hence, production and decay of heavy quarkonium can occur within NRQCD only through hadronic and electromagnetic transitions.
However, the time scales associated with these processes  are much larger
than the time scale associated with the hadronization of $Q \bar Q$ into heavy quarkonium, which is typically of order $1/\Lambda_{\rm QCD}$. 
Therefore, to a good approximation, we can neglect transition processes in describing the hadronization process within NRQCD.
We conclude that the operator ${\cal P}_{{\cal Q}(\bm{P})}$, which counts the number of heavy quarkonia, is conserved.
Since it commutes with the NRQCD Hamiltonian, ${\cal P}_{{\cal Q}(\bm{P})}$ and $H_{\rm NRQCD}$ can be diagonalized simultaneously.
A simultaneous eigenstate of ${\cal P}_{{\cal Q}(\bm{P})}$ and $H_{\rm NRQCD}$ has the form
\begin{equation} 
\label{quarkonium_state} 
| {\cal Q} (n,\bm{P}) \rangle = \int d^3x_1 d^3x_2 \, \phi_{{\cal Q}(n,\bm{P})}(\bm{x}_1,\bm{x}_2) \, |\underline{\rm n}; \bm{x}_1, \bm{x}_2 \rangle.
\end{equation} 
The states $|\underline{\rm n}; \bm{x}_1, \bm{x}_2 \rangle$ have been introduced above. 
They may transform as color singlets or color octets.
Only color-singlet states project on $\langle \Omega |\chi^\dag {\cal K}_N \psi$ and
only color-octet states project on $\langle \Omega | \chi^\dag {\cal K}_N T^a \psi \Phi_\ell^{\dag ab} (0)$.
In $|\underline{\rm n}; \bm{x}_1, \bm{x}_2 \rangle$ we have dropped the color label because the spectrum, $E_n(\bm{x}_1, \bm{x}_2; \bm{\nabla}_1, \bm{\nabla}_2)$, does not depend on it.
Color labels are implicit for color-octet states.
The functions $\phi_{{\cal Q}(n,\bm{P})}(\bm{x}_1,\bm{x}_2)$ are eigenfunctions of $E_n(\bm{x}_1, \bm{x}_2; \bm{\nabla}_1, \bm{\nabla}_2)$.
They depend on the heavy quark and antiquark locations, on the quarkonium ${\cal Q}$ that moves with momentum $\bm{P}$,
and on the light degrees of freedom through the quantum number $n$.
For $n=0$ the state in eq.~\eqref{quarkonium_state} is just a heavy quarkonium state~\cite{Brambilla:2002nu}.
We take the state $|{\cal Q}(n,\bm{P})\rangle$ to be nonrelativistically normalized.  

Infinitely massive quark-antiquark pairs must be in a color-singlet state at the production point $\bm{x}_1 = \bm{x}_2$ in order to overlap with a state containing a quarkonium.  
This requirement limits the number of states that contribute to ${\cal P}_{{\cal Q}(\bm{P})}$.
In order to contribute to ${\cal P}_{{\cal Q}(\bm{P})}$, states must have a static component $|\underline{\rm n}; \bm{x}_1, \bm{x}_2 \rangle^{(0)}$
that fulfills 
\begin{equation} 
\delta^{(3)}(\bm{r})|\underline{\rm n}; \bm{x}_1, \bm{x}_2 \rangle^{(0)} =
\delta^{(3)}(\bm{r}) \frac{1}{N_c} \, {\rm Tr} \{ \psi^\dag (\bm{x}_1) \chi(\bm{x}_2) \} \, {\rm Tr} \{ |n; \bm{x}_1, \bm{x}_2 \rangle^{(0)} \},
\label{Sset}
\end{equation} 
where ${\rm Tr}$ is the trace over the color indices.
Property \eqref{Sset} will turn out to be crucial in evaluating the LDMEs, which, involving the local operators 
$\chi^\dag {\cal K}_N \psi$, $\psi^\dag {\cal K}'_N \chi$,  $\chi^\dag {\cal K}_N T^a \psi$ and $\psi^\dag {\cal K}'_N T^c \chi$,
require the quark and antiquark to be located at the same point.
We denote by $\mathbb S$ the subset of eigenstates that fulfill the requirement \eqref{Sset}.\footnote{
\label{footindex2}
  The condition \eqref{Sset} refers to the state $|\underline{\rm n}; \bm{x}_1, \bm{x}_2 \rangle^{(0)}$.
  The state  $|n; \bm{x}_1, \bm{x}_2 \rangle^{(0)}$ belongs to the subset $\mathbb S$ if it fulfills the condition (color indices are explicit)
  $$
  \delta^{(3)}(\bm{r})|n; \bm{x}_1, \bm{x}_2; i,j \rangle^{(0)} =
  \delta^{(3)}(\bm{r}) \frac{\delta_{ij}}{N_c} \, |n; \bm{x}_1, \bm{x}_2; k,k \rangle^{(0)}.
  $$
}
Then, the explicit expression of the projector ${\cal P}_{{\cal Q}(\bm{P})}$ is 
\begin{equation} 
\label{eq:projector_result1} 
{\cal P}_{{\cal Q}(\bm{P})} = \sum_{n \in {\mathbb S}} | {\cal Q} (n,\bm{P}) \rangle \langle {\cal Q} (n,\bm{P}) |\,.
\end{equation} 
It implies that $\langle \underline{\rm n}; \bm{x}_1, \bm{x}_2 | {\cal P}_{{\cal Q}(\bm{P})} | \underline{\rm k}; \bm{x}_1', \bm{x}_2' \rangle = 0$ for $n$ or $k \notin {\mathbb S}$
and for $n \neq k$, while $\langle \underline{\rm n}; \bm{x}_1, \bm{x}_2|$ ${\cal P}_{{\cal Q}(\bm{P})} | \underline{\rm n}; \bm{x}_1', \bm{x}_2' \rangle$ 
is completely determined by the function $\phi_{{\cal Q}(n,\bm{P})}(\bm{x}_1,\bm{x}_2)$ when $n \in {\mathbb S}$.  
We will now show how to compute these functions at leading order in $v$.  

Under the assumption $mv^2 \ll \Lambda_{\rm QCD}$, the dynamics of the degrees of freedom scaling with the energy scale $mv^2$ is described by the Hamiltonian of strongly coupled pNRQCD.
Strongly coupled pNRQCD follows from NRQCD by integrating out degrees of freedom carrying momentum or energy of order $mv$ or $\Lambda_{\rm QCD}$~\cite{Brambilla:1999xf,Brambilla:2000gk,Pineda:2000sz}.
It is assumed that its degrees of freedom consist of color-singlet states made of a heavy quark-antiquark pair and light degrees of freedom.
If the effect of dynamical light quarks is neglected (the coupling of light
mesons to quarkonia below threshold is a subleading effect, suppressed by powers of momenta of the light mesons, which is of order  $mv^2$),
then the dynamical degrees of freedom of pNRQCD consist of quarkonia and quarkonium exotica~\cite{Brambilla:2001xy,Brambilla:2002nu,Brambilla:2020xod}.
These degrees of freedom are represented by fields $S_n$ annihilating a color-singlet heavy quark-antiquark pair with light degrees of freedom in a state $n$.
If we assume the simplest scenario that the energy levels for different $n$ are separated by a gap  of order $\Lambda_{\rm QCD}$,
or that the mixing between fields with different $n$ may be neglected, then we can isolate the dynamics of one single field $S_n$.
The strongly coupled pNRQCD Hamiltonian that describes this dynamics is  
\begin{equation} 
H_{\rm pNRQCD} = \int d^3x_1 \, d^3x_2 \;  S_n^\dag \,h_n (\bm{x}_1, \bm{x}_2; \bm{\nabla}_1, \bm{\nabla}_2) \,S_n.
\end{equation} 
The function $h_n (\bm{x}_1, \bm{x}_2; \bm{\nabla}_1, \bm{\nabla}_2) \delta^{(3)} (\bm{x}_1 - \bm{x}_1') \delta^{(3)} (\bm{x}_2 - \bm{x}_2')$ is determined
by matching to the NRQCD matrix element $\langle \underline{\rm n}; \bm{x}_1, \bm{x}_2 | H_{\rm NRQCD}$ $|\underline{\rm n}; \bm{x}_1',  \bm{x}_2'\rangle$ of eq.~\eqref{braHket}.
At leading order in $v$, $h_n (\bm{x}_1, \bm{x}_2; \bm{\nabla}_1, \bm{\nabla}_2) $ is given by 
\begin{equation} 
\label{eq:schroedinger} 
h_n(\bm{x}_1, \bm{x}_2; \bm{\nabla}_1, \bm{\nabla}_2) = - \frac{\bm{\nabla}_1^2}{2 m} - \frac{\bm{\nabla}_2^2}{2 m} + V^{(0;n)} (\bm{x}_1, \bm{x}_2), 
\end{equation} 
where $- \bm{\nabla}_1^2/(2 m) - \bm{\nabla}_2^2/(2 m)$ is the kinetic energy of the heavy quark-antiquark pair and $V^{(0;n)}(\bm{x}_1, \bm{x}_2)$ is the static potential.
The matching fixes the static potential to be the energy eigenvalue $E_n^{(0)}(\bm{x}_1, \bm{x}_2)$ of $H_{\rm NRQCD}^{(0)}$.  
As a consequence of the matching, the functions $\phi_{{\cal Q}(n,\bm{P})}(\bm{x}_1,\bm{x}_2)$ are also eigenfunctions of $h_n$.

For $n=0$, $V^{(0;n=0)}$ is the QCD static potential, which can be determined from the vacuum expectation value of a static Wilson loop.
In particular, it holds that 
\begin{equation} 
\label{eq:WL} 
V^{(0;n=0)} (\bm{x}_1, \bm{x}_2) = E_{n=0}^{(0)}(\bm{x}_1, \bm{x}_2) = \lim_{T \to \infty} \frac{i}{T} \log \langle \Omega | W_{r \times T} | \Omega \rangle, 
\end{equation} 
where $W_{r \times T}$ is a rectangular Wilson loop of spatial and temporal lengths $r=|\bm{x}_1-\bm{x}_2|$ and $T$, respectively.  
Then, the function $\phi_{{\cal Q}(n=0,\bm{P})}(\bm{x}_1,\bm{x}_2)$ at leading order in $v$ is given by 
\begin{equation} 
\label{eq:WF0} 
\phi_{{\cal Q}(n=0,\bm{P})}(\bm{x}_1,\bm{x}_2) \approx e^{i \bm{P} \cdot (\bm{x}_1+\bm{x}_2)/2} \phi^{(0)}_{{\cal Q}} (\bm{x}_1-\bm{x}_2), 
\end{equation} 
where $e^{i \bm{P} \cdot (\bm{x}_1+\bm{x}_2)/2}$ is a plane wave encoding the center-of-mass motion,
and $\phi^{(0)}_{{\cal Q}} (\bm{x}_1-\bm{x}_2)$ is the quarkonium wavefunction solution of the Schr\"odinger equation with Hamiltonian $h_{n=0}$.
Because of translational invariance, the static potential depends only on $\bm{r} = \bm{x}_1-\bm{x}_2$.

For $n \in {\mathbb S}$ and $n\neq 0$, the static potential $V^{(0;n)}$ can be obtained from the vacuum expectation value of a static Wilson loop
whose initial and final states select the excitation $n$ of the static sources.
While lattice QCD determinations of $V^{(0;n)}$ for $n \in {\mathbb S}$ and $n \neq 0$ are not available yet,
we expect the disconnected gluon fields to produce mainly a constant shift to the potentials,
possibly in the form of a glueball mass, and do not significantly affect the slopes.
This is also supported by the large $N_c$ limit,
where the vacuum expectation value of a Wilson loop with additional disconnected gluon fields factorizes into the vacuum expectation value of the Wilson loop
times the vacuum expectation value of the additional gluon fields up to corrections of order $1/N_c^2$~\cite{Makeenko:1979pb, Witten:1979pi}.  
If the slopes of the static potentials are the same for all $n \in {\mathbb S}$,
then the wavefunctions $\phi_{{\cal Q}(n,\bm{P})}(\bm{x}_1,\bm{x}_2)$ are given by eq.~\eqref{eq:WF0}, and are independent of $n$.
Hence, we take the approximation 
\begin{equation} 
\label{eq:WFn} 
\phi_{{\cal Q}(n,\bm{P})}(\bm{x}_1,\bm{x}_2)  \approx e^{i \bm{P} \cdot (\bm{x}_1+\bm{x}_2)/2} \phi^{(0)}_{{\cal Q}} (\bm{x}_1-\bm{x}_2), 
\end{equation} 
making an error that is at most of order $1/N_c^2$ and of order $v^2$.  

We can now compute the matrix elements of ${\cal P}_{{\cal Q}(\bm{P}=\bm{0})}$ on the states $|\underline{\rm n}; \bm{x}_1, \bm{x}_2 \rangle$.
From eqs.~\eqref{eq:projector_result1}, \eqref{quarkonium_state}, and \eqref{eq:WFn}, we obtain 
\begin{align} 
\langle \underline{\rm n}; \bm{x}_1, \bm{x}_2 | {\cal P}_{{\cal Q}(\bm{P}=\bm{0})} | \underline{\rm n}; \bm{x}_1', \bm{x}_2' \rangle 
&= \phi_{{\cal Q} (n,\bm{P}=\bm{0})} (\bm{x}_1,\bm{x}_2) \phi_{{\cal Q} (n,\bm{P}=\bm{0})}^* (\bm{x}_1',\bm{x}_2') 
\nonumber\\ 
\label{eq:poperator_result} 
&= \phi^{(0)}_{{\cal Q}} (\bm{x}_1-\bm{x}_2) \phi^{(0)\,*}_{{\cal Q}} (\bm{x}_1'-\bm{x}_2') + O\left(\frac{1}{N_c^2},v^2\right).  
\end{align} 
This leads to the pNRQCD expression for the operator ${\cal P}_{{\cal Q}(\bm{P}=\bm{0})}$:
\begin{align}
  {\cal P}_{{\cal Q}(\bm{P}=\bm{0})}  =&  \sum_{n \in {\mathbb S}}   \int d^3x_1 d^3x_2 d^3x'_1 d^3x'_2 \,
 |\underline{\rm n}; \bm{x}_1, \bm{x}_2 \rangle \, \phi^{(0)}_{{\cal Q}} (\bm{x}_1-\bm{x}_2) \phi^{(0)\,*}_{{\cal Q}} (\bm{x}_1'-\bm{x}_2')\, \langle \underline{\rm n}; \bm{x}'_1, \bm{x}'_2 |
  \nonumber\\
& +  O\left(\frac{1}{N_c^2},v^2\right).
  \label{eq:poperator_result_bis}
\end{align}

\subsection{Matching of the LDMEs} 
Equation~\eqref{eq:poperator_result_bis} implies that a LDME can be matched into the following pNRQCD expression
\begin{align} 
\label{eq:ldme_pnrqcd} 
\langle \Omega | {\cal O}^{\cal Q}(N) | \Omega \rangle 
= & 
\frac{1}{\langle \bm{P}=\bm{0} | \bm{P} = \bm{0} \rangle} \int d^3x_1 d^3x_2 d^3x'_1 d^3x'_2 \, \phi^{(0)}_{{\cal Q}} (\bm{x}_1-\bm{x}_2) 
\nonumber \\ & 
\times \left[ - V_{{\cal O}(N)} (\bm{x}_1, \bm{x}_2;\bm{\nabla}_1, \bm{\nabla}_2) 
\delta^{(3)} (\bm{x}_1-\bm{x}'_1) \delta^{(3)} (\bm{x}_2-\bm{x}'_2) \right] \phi^{(0)\,*}_{{\cal Q}} (\bm{x}_1'-\bm{x}_2'), 
\end{align} 
where $V_{{\cal O}(N)}$ is a contact term.
For the color-singlet operator ${\cal O}^{\cal Q}(N_{\textrm{color singlet}})$ given in eq.~\eqref{eq:cs_ldme}, the contact term reads 
\begin{align} 
\label{eq:cs_contact} 
& \sum_{n \in {\mathbb S}} \int d^3x \, \langle \Omega | \left( \chi^\dag {\cal K}_N \psi \right) (\bm{x}) | \underline{\rm n}; \bm{x}_1, \bm{x}_2 \rangle 
\langle \underline{\rm n}; \bm{x}'_1, \bm{x}'_2 | \left( \psi^\dag {\cal K}'_N \chi \right) (\bm{x}) | \Omega \rangle 
\nonumber \\ 
& \hspace{4cm}
    = - V_{{\cal O}(N)} (\bm{x}_1, \bm{x}_2;\bm{\nabla}_1, \bm{\nabla}_2) \, \delta^{(3)} (\bm{x}_1-\bm{x}'_1) \delta^{(3)} (\bm{x}_2-\bm{x}'_2).  
\end{align} 
The space integration is compensated by the denominator $\displaystyle \langle \bm{P}=\bm{0}| \bm{P} = \bm{0} \rangle \displaystyle = \int d^3x$ in eq.~\eqref{eq:ldme_pnrqcd}.
Similarly, for the color-octet operator ${\cal O}^{\cal Q}(N_{\textrm{color octet}})$ given in eq.~\eqref{eq:co_ldme}, the contact term reads
\begin{align} 
\label{eq:co_contact} 
& \sum_{n \in {\mathbb S}} \int d^3x \langle \Omega | \left( \chi^\dag {\cal K}_N T^a \psi \right) (\bm{x}) \Phi_\ell^{\dag ab} (0,\bm{x}) | \underline{\rm n}; \bm{x}_1, \bm{x}_2 \rangle 
\langle \underline{\rm n}; \bm{x}'_1, \bm{x}'_2 | \Phi_\ell^{bc}(0,\bm{x}) \left( \psi^\dag {\cal K}'_N T^c \chi \right) (\bm{x}) | \Omega \rangle 
\nonumber \\ 
&\hspace{4cm}
    = - V_{{\cal O}(N)} (\bm{x}_1, \bm{x}_2;\bm{\nabla}_1, \bm{\nabla}_2) \delta^{(3)} (\bm{x}_1-\bm{x}'_1) \delta^{(3)} (\bm{x}_2-\bm{x}'_2).  
\end{align} 

The contact terms in eqs.~\eqref{eq:cs_contact} and \eqref{eq:co_contact} can be computed by substituting the states 
$|\underline{\rm n}; \bm{x}_1, \bm{x}_2 \rangle$ with their quantum-mechanical expansion in powers of $1/m$, see eq.~\eqref{QMPT2},   
and by making explicit the heavy quark and antiquark fields using eq.~\eqref{nbarvsn}.
The heavy quark and antiquark fields are then removed by using Wick's theorem.  
The calculation of the contact terms is done in a similar way as the computation of the decay matrix elements in strongly coupled pNRQCD, 
except that the intermediate states are not just a quarkonium state, but include all states belonging to ${\mathbb S}$ and, 
in the calculation of the color-octet production matrix elements, the gauge-completion Wilson lines must be included.
As a result, the contact term $V_{{\cal O}(N)}$ is proportional to the delta function $\delta^{(3)}(\bm{r})$ or to derivatives of it.
Furthermore, it can involve matrix elements of gluon fields that may be eventually expressed in terms of temporal correlators of gluon fields 
by using the techniques developed in refs.~\cite{Brambilla:2001xy,Brambilla:2002nu}.
Finally, the LDMEs for inclusive heavy quarkonium production are computed from eq.~\eqref{eq:ldme_pnrqcd}.  
Because of the form of the contact terms, they turn out to depend on the quarkonium wavefunctions at the origin or its derivatives, and on vacuum expectation values of gluon fields.

The contact term for the color-octet matrix element in eq.~\eqref{eq:co_contact} can in principle depend on the direction $\ell$ of the gauge-completion Wilson lines $\Phi_\ell$.
The dependence on the direction $\ell$ disappears in the LDMEs if the quarkonium ${\cal Q}$ has zero angular momentum,
because the product of wavefunctions $\phi_{\cal Q}^{(0)} (\bm{x}_1-\bm{x}_2) \phi_{\cal Q}^{(0)\,*} (\bm{x}_1'-\bm{x}_2')$ in the integrand of eq.~\eqref{eq:ldme_pnrqcd} is isotropic.  
The $\ell$ dependence also disappears in color-octet matrix elements for production of a quarkonium with nonzero angular momentum, if we sum over the quarkonium polarizations.
The disappearance of the $\ell$ dependence is a necessary condition for the NRQCD factorization to hold~\cite{Nayak:2005rw, Nayak:2005rt}. 
Therefore, we can already state, on general grounds, that the calculation in pNRQCD of the LDMEs
will support their universality in the case of unpolarized and polarization-summed production cross sections of heavy quarkonia.  

In order to be consistent with the perturbative factorization of eq.~\eqref{eq:NRQCDfac}, the right-hand side of eq.~\eqref{eq:ldme_pnrqcd}, 
when computed in perturbative QCD, must have the same IR divergences as the NRQCD counterpart,
which is obtained by computing in perturbative QCD the vacuum expectation values of the operators in eqs.~\eqref{eq:cs_ldme} and \eqref{eq:co_ldme}.
In section \ref{sec:consistency}, we will confirm this agreement for the color-octet matrix elements that appear in the production cross sections of $P$-wave quarkonia.

\section{\boldmath Theory of inclusive production of $P$-wave quarkonia} 
\label{sec:pwave-th}

\subsection[$P$-wave LDMEs in pNRQCD]{\boldmath $P$-wave LDMEs in pNRQCD} 
Based on the formalism developed in the previous section, we compute here the LDMEs that appear in production cross sections of $P$-wave heavy quarkonia,
which include $h_Q$ and $\chi_{QJ}$, where $Q=c$ or $b$ and $J=0$, 1, and 2.  
The NRQCD factorization formula for the inclusive production cross sections of $P$-wave quarkonia at leading order in $v$ read 
\begin{subequations} 
\label{eq:fac_pwave} 
\begin{align} 
  \sigma_{h_Q+X} &= \sigma_{Q \bar Q({}^1P_1^{[1]})} \langle \Omega | {\cal O}^{h_Q}({}^1P_1^{[1]}) | \Omega \rangle
                   + \sigma_{Q \bar Q({}^1S_0^{[8]})} \langle \Omega | {\cal O}^{h_Q}({}^1S_0^{[8]}) | \Omega \rangle, 
\\ 
  \sigma_{\chi_{QJ}+X} &= \sigma_{Q \bar Q({}^3P_J^{[1]})} \langle \Omega | {\cal O}^{\chi_{QJ}}({}^3P_J^{[1]}) | \Omega \rangle
                         + \sigma_{Q \bar Q({}^3S_1^{[8]})} \langle \Omega | {\cal O}^{\chi_{QJ}}({}^3S_1^{[8]}) | \Omega \rangle.
\end{align} 
\end{subequations} 
The operators are
\begin{subequations} 
\begin{align} 
\label{eq:pwave_operators} 
{\cal O}^{h_Q}({}^1P_1^{[1]}) &= 
\sum_{\lambda} \chi^\dag \left(-\frac{i}{2} \overleftrightarrow{D}^i \right) \psi \, 
{\cal P}_{{h_Q}(\lambda,\bm{P}=\bm{0})} \, \psi^\dag \left(-\frac{i}{2} \overleftrightarrow{D}^i \right) \chi, 
\\ 
{\cal O}^{h_Q}({}^1S_0^{[8]}) &= 
\sum_{\lambda} \chi^\dag T^a \psi \Phi_\ell^{\dag ab} (0) \, {\cal P}_{{h_Q}(\lambda, \bm{P}=\bm{0})} \, \Phi_\ell^{bc} (0) \psi^\dag T^c \chi, 
\\ 
{\cal O}^{\chi_{Q0}}({}^3P_0^{[1]}) &= 
\frac{1}{3} \chi^\dag \left(-\frac{i}{2} \overleftrightarrow{\bm{D}} \cdot \bm{\sigma} \right) \psi
\, {\cal P}_{\chi_{Q0}(\bm{P}=\bm{0})} \, \psi^\dag \left(-\frac{i}{2} \overleftrightarrow{\bm{D}} \cdot \bm{\sigma} \right) \chi, 
\\ 
{\cal O}^{\chi_{Q1}}({}^3P_1^{[1]}) &= 
\sum_{\lambda} \frac{1}{2} \chi^\dag \left(-\frac{i}{2} \overleftrightarrow{\bm{D}} \times \bm{\sigma} \right)^i \psi \, {\cal P}_{\chi_{Q1}(\lambda, \bm{P}=\bm{0})} \, 
\psi^\dag \left(-\frac{i}{2} \overleftrightarrow{\bm{D}} \times \bm{\sigma} \right)^i \chi, 
\\ 
{\cal O}^{\chi_{Q2}}({}^3P_2^{[1]}) &= 
\sum_{\lambda} \chi^\dag \left(-\frac{i}{2} \overleftrightarrow{{D}}^{(i} {\sigma}^{j)} \right) \psi \, {\cal P}_{\chi_{Q2}(\lambda, \bm{P}=\bm{0})} \,
\psi^\dag \left(-\frac{i}{2} \overleftrightarrow{{D}}^{(i} {\sigma}^{j)} \right) \chi, 
\\ 
{\cal O}^{\chi_{QJ}}({}^3S_1^{[8]}) &= 
\sum_{\lambda} \chi^\dag \sigma^i T^a \psi \Phi_\ell^{\dag ab} (0) \, {\cal P}_{{\chi_{QJ}}(\lambda, \bm{P}=\bm{0})} \, \Phi_\ell^{bc} (0) \psi^\dag \sigma^i T^c \chi,  
\end{align} 
\end{subequations} 
where we have used the notation $\displaystyle A^{(i\,j)} = \frac{A^{ij} + A^{ji}}{2} - \frac{\delta^{ij}}{3}A^{kk}$.
The parameter $\lambda$ is the polarization of the quarkonium with nonzero angular momentum.
We sum over all polarizations $\lambda$ when computing polarization-summed cross sections,
while we do not sum when computing polarized cross sections.
In this section, we restrict to the case of polarization-summed cross sections.

Since we neglect transition processes between heavy quarkonium states in our treatment of the LDMEs,
the inclusive production cross sections that we compute from the NRQCD factorization formula include only the ``direct'' production rates,
where feeddown contributions that come from decays of higher quarkonium states are neglected. 
The feeddown contributions to inclusive quarkonium production cross sections can be included by adding direct production cross sections of higher quarkonium states,
multiplied by the branching ratios into the measured quarkonium (see, for instance, section~\ref{sec:chibprod}). 

We first match the contact term for the operator ${\cal O}^{h_Q}({}^1P_1^{[1]})$.
At leading order in the quantum-mechanical perturbation theory, it holds that  
\begin{align} 
& 
- V_{{\cal O} ({}^1P_1^{[1]})} \delta^{(3)} (\bm{x}_1 - \bm{x}_1') \delta^{(3)} (\bm{x}_2 - \bm{x}_2') 
\nonumber \\ 
&\hspace{3cm}= 
                \sum_{n \in {\mathbb S}} \int d^3x \,
                \langle \Omega | \left[ \chi^\dag \left(-\frac{i}{2} \overleftrightarrow{D}^i \right) \psi \right]\!\! (\bm{x})\; | \underline{\rm n}; \bm{x}_1, \bm{x}_2 \rangle ^{(0)} 
\nonumber \\ & \hspace{30ex} \times 
{}^{(0)} \langle \underline{\rm n}; \bm{x}'_1, \bm{x}'_2 | \left[ \psi^\dag \left(-\frac{i}{2} \overleftrightarrow{D}^i \right) \chi \right]\!\! (\bm{x})\; | \Omega \rangle 
\nonumber \\ 
&\hspace{3cm}= 
\sum_{n \in {\mathbb S}} \int d^3x \, 
\langle \Omega | \delta^{(3)} (\bm{x}_2 - \bm{x}) \left(-\frac{i}{2} \overleftrightarrow{D}^i(\bm{x}) \right) \delta^{(3)} (\bm{x}_1 - \bm{x}) | n; \bm{x}_1, \bm{x}_2 \rangle ^{(0)} 
\nonumber \\ & \hspace{30ex} \times 
{}^{(0)}\langle n; \bm{x}'_1, \bm{x}'_2 | \delta^{(3)} (\bm{x}'_1 - \bm{x}) \left(-\frac{i}{2} \overleftrightarrow{D}^i(\bm{x}) \right) \delta^{(3)} (\bm{x}'_2 - \bm{x}) | \Omega \rangle 
\nonumber \\ 
&\hspace{3cm}\underset{\textrm{$P$-wave}}{=} 
   -N_c \nabla_{\bm{r}}^i \delta^{(3)} (\bm{r}) \nabla_{\bm{r}}^i \delta^{(3)} (\bm{x}_1 - \bm{x}'_1) \delta^{(3)} (\bm{x}_2 - \bm{x}'_2),
\label{eq:contact_hqsinglet} 
\end{align} 
where $\bm{r}= \bm{x}_1 - \bm{x}_2$ and $\bm{\nabla}_{\bm{r}} = (\bm{\nabla}_1 - \bm{\nabla}_2)/2$.
In the last equality, we keep only  terms that give nonzero contributions when  inserted in eq.~(\ref{eq:ldme_pnrqcd}) to compute LDMEs for $P$-wave quarkonium states. 
Because the $P$-wave wavefunctions $\phi^{(0)}_{{}^1P_1} (\bm{r})$ vanish at the origin, only terms in $V_{{\cal O} ({}^1P_1^{[1]})}$ that contain two derivatives can make
nonvanishing contributions to the right-hand side of eq.~\eqref{eq:ldme_pnrqcd}.
We get terms proportional to derivatives only if the intermediate state is taken to be $n=0$,
an observation that follows from eq.~\eqref{0vsOmega}, the canonical commutation    relations and symmetry considerations~\cite{Brambilla:2000gk,Brambilla:2004jw};
the $n=0$ state eventually reduces to the vacuum state so that $\langle \Omega| \bm{D}|\Omega \rangle = \bm{\nabla}$. 
From \eqref{eq:contact_hqsinglet} we get 
\begin{align} 
\left. - V_{{\cal O} ({}^1P_1^{[1]})} \right|_{\textrm{$P$-wave}} = -N_c \nabla_{\bm{r}}^i \delta^{(3)} (\bm{r}) \nabla_{\bm{r}}^i,
\label{eq:contact_hqsinglet2} 
\end{align} 
which implies at leading order in $v$
\begin{equation} 
\label{eq:hqsinglet} 
\langle \Omega | {\cal O}^{h_Q}({}^1P_1^{[1]}) | \Omega \rangle = 3 \times \frac{3 N_c}{2\pi} |R^{(0)}{}'(0)|^2, 
\end{equation} 
where the radial wavefunction $R^{(0)} (r)$ is defined through the relation $\phi^{(0)}_{{}^1P_1} (\bm{r}) = R^{(0)} (r)$ $Y_1^\lambda (\hat{\bm{r}})$, 
$\lambda$ being the polarization of the ${}^1P_1$ state and $Y_1^\lambda (\hat{\bm{r}})$ the spherical harmonics for $P$-wave states.  
The factor $3$ in eq.~\eqref{eq:hqsinglet} comes from the sum over the 3 polarizations of the $h_Q$,
while the factor $3/(2 \pi)$ comes from the trace over the heavy quark spin times the normalization $\sum_\lambda Y_1^\lambda (\hat{\bm{r}}) Y_1^\lambda{}^* (\hat{\bm{r}}) = 3/(4 \pi)$.  
At leading order in $v$, the radial wavefunction is independent of the polarization $\lambda$.  
Equation \eqref{eq:hqsinglet} reproduces the well-known result obtained in the vacuum-saturation approximation~\cite{Bodwin:1994jh}.  

In the following paragraphs, we match the contact term for the color-octet operator ${\cal O}^{h_Q}({}^1S_0^{[8]})$.
At leading order in the quantum-mechanical perturbation theory, we have 
\begin{align} 
- V_{{\cal O} ({}^1S_0^{[8]})} \delta^{(3)} (\bm{x}_1 - \bm{x}_1') \delta^{(3)} (\bm{x}_2 - \bm{x}_2') 
&= 
\sum_{n \in {\mathbb S}} \int d^3x \, \langle \Omega | \left( \chi^\dag T^a \psi \right) (\bm{x}) \Phi_\ell^{\dag ab}(0,\bm{x}) | \underline{\rm n}; \bm{x}_1, \bm{x}_2 \rangle ^{(0)} 
\nonumber \\ & \hspace{6ex} \times 
{}^{(0)} \langle \underline{\rm n}; \bm{x}'_1, \bm{x}'_2 | \Phi_\ell^{bc}(0,\bm{x}) \left( \psi^\dag T^c \chi \right) (\bm{x}) | \Omega \rangle.  
\end{align} 
This expression leads to a vanishing contribution to the LDME  once inserted in the right-hand side of eq.~\eqref{eq:ldme_pnrqcd}.
The reason is twofold: 
first, there are no derivatives acting on the $P$-wave wavefunctions
and second, the heavy quark-antiquark pair in the state $|\underline{\rm n}; \bm{x}_1, \bm{x}_2 \rangle ^{(0)}$ behaves like a color singlet at $\bm{r}=0$ (see eq.~\eqref{Sset}),
which leads to a trace over an SU(3) generator after application of Wick's theorem.  

In order to obtain a nonvanishing contribution to the octet LDME, 
we need to include $1/m$ corrections to the states
$|\underline{\rm n}; \bm{x}_1, \bm{x}_2\rangle$ appearing in the left-hand side of eq.~\eqref{eq:co_contact}.
These corrections have been first derived in refs.~\cite{Brambilla:2000gk, Brambilla:2002nu}.  
Moreover, because the color-octet operator ${\cal O}^{h_Q}({}^1S_0^{[8]})$ does not contain derivatives, 
we need to keep in $|\underline{\rm n};\bm{x}_1,\bm{x}_2\rangle^{(1)}$ only the part that does contain derivatives.
We denote this part with $|\underline{\rm n};\bm{x}_1,\bm{x}_2\rangle^{(1)}_{\textrm{$P$-wave}}$:
\begin{align} 
\label{eq:state_pwave} 
| \underline{\rm n}; \bm{x}_1, \bm{x}_2 \rangle^{(1)}_{P\text{-wave}} =& - \sum_{k \neq n} |\underline{\rm k}; \bm{x}_1, \bm{x}_2\rangle^{(0)} 
\nonumber\\
& \hspace{-27mm}
              \times
\left[\frac{{}^{(0)}\langle k; \bm{x}_1, \bm{x}_2 | g\bm{E}_{1}|n; \bm{x}_1, \bm{x}_2 \rangle^{(0)}}{(E^{(0)}_{n}(\bm{x}_1,\bm{x}_2)-E^{(0)}_{k}(\bm{x}_1,\bm{x}_2))^2}\cdot \overleftarrow{\bm{\nabla}}_{1} 
- \frac{{}^{(0)}\langle k; \bm{x}_1, \bm{x}_2 | g\bm{E}_{2}^T|n; \bm{x}_1, \bm{x}_2 \rangle^{(0)}}{(E^{(0)}_{n}(\bm{x}_1,\bm{x}_2)-E^{(0)}_{k}(\bm{x}_1,\bm{x}_2))^2}\cdot \overleftarrow{\bm{\nabla}}_{2}\right],
\end{align} 
where the fields $\bm{E}_1$ and $\bm{E}_2$ are the chromoelectric fields computed at the positions $\bm{x}_1$ and $\bm{x}_2$, respectively.  
The contribution of  $|\underline{\rm n};\bm{x}_1,\bm{x}_2\rangle^{(1)}_{\textrm{$P$-wave}}$ to the contact term for the color-octet operator ${\cal O}^{h_Q}({}^1S_0^{[8]})$ is
\begin{align} 
& \hspace{-3ex} 
- V_{{\cal O} ({}^1S_0^{[8]})} \delta^{(3)} (\bm{x}_1 - \bm{x}_1') \delta^{(3)} (\bm{x}_2 - \bm{x}_2') \Big|_{\textrm{$P$-wave}} 
\nonumber \\ 
  &= \frac{1}{m^2} \sum_{n \in {\mathbb S}} \int d^3x\,
    \langle \Omega | \left( \chi^\dag T^a \psi \right) \!(\bm{x}) \; \Phi_\ell^{\dag ab}(0,\bm{x}) | \underline{\rm n}; \bm{x}_1, \bm{x}_2 \rangle^{(1)}_{\textrm{$P$-wave}} 
\nonumber \\ & \hspace{8ex} \times 
               {}_{\textrm{$P$-wave}}{\hspace{-10pt}}^{(1)} \langle \underline{\rm n}; \bm{x}'_1, \bm{x}'_2 | \Phi_\ell^{bc}(0,\bm{x}) \; \left( \psi^\dag T^c \chi \right) \!(\bm{x}) | \Omega \rangle 
\nonumber \\ 
&= 
                \frac{1}{m^2} \sum_{n \in {\mathbb S}} \sum_{p \neq n} \sum_{k \neq n} \int d^3x\,
                \langle \Omega | \left( \chi^\dag T^a \psi \right)\! (\bm{x}) \; \Phi_\ell^{\dag ab}(0,\bm{x}) | \underline{\rm p}; \bm{x}_1, \bm{x}_2 \rangle ^{(0)} 
\nonumber \\ 
& \hspace{8ex} \times 
\left( \frac{{}^{(0)} \langle p ; \bm{x}_1, \bm{x}_2 | g \bm{E}_1 | n ; \bm{x}_1, \bm{x}_2 \rangle^{(0)}}{[E_n^{(0)}(\bm{x}_1,\bm{x}_2)-E_p^{(0)}(\bm{x}_1,\bm{x}_2)]^2} \cdot \overleftarrow{\bm{\nabla}}_1 
- \frac{{}^{(0)} \langle p ; \bm{x}_1, \bm{x}_2| g \bm{E}_2^T | n ; \bm{x}_1, \bm{x}_2 \rangle^{(0)}}{[E_n^{(0)}(\bm{x}_1,\bm{x}_2)-E_p^{(0)}(\bm{x}_1,\bm{x}_2)]^2} \cdot \overleftarrow{\bm{\nabla}}_2 \right) 
\nonumber \\ 
&\hspace{8ex} \times
 \left( \bm{\nabla}_1'  \cdot \frac{{}^{(0)} \langle n; \bm{x}'_1, \bm{x}'_2 | g \bm{E}_1 | k; \bm{x}'_1, \bm{x}'_2 \rangle^{(0)}}{[E_n^{(0)}(\bm{x}_1',\bm{x}_2')-E_k^{(0)}(\bm{x}_1',\bm{x}_2')]^2}
-\bm{\nabla}_2' \cdot \frac{{}^{(0)} \langle n; \bm{x}'_1, \bm{x}'_2 | g \bm{E}_2^T | k; \bm{x}'_1, \bm{x}'_2 \rangle^{(0)}}{[E_n^{(0)}(\bm{x}_1',\bm{x}_2')-E_k^{(0)}(\bm{x}_1',\bm{x}_2')]^2} \right) 
\nonumber \\ 
&\hspace{8ex} \times 
{}^{(0)} \langle \underline{\rm k}; \bm{x}'_1, \bm{x}'_2 | \Phi_\ell^{bc}(0,\bm{x})\; \left( \psi^\dag T^c \chi \right)\! (\bm{x}) | \Omega \rangle.  
\label{eq:contact_hqoctet2} 
\end{align} 
The matrix elements still containing heavy quark and antiquark fields in eq.~\eqref{eq:contact_hqoctet2} can be evaluated as follows: 
\begin{align} 
& 
\sum_{p \neq n}\langle \Omega | \left( \chi^\dag T^a \psi \right) \! (\bm{x}) \; \Phi_\ell^{\dag ab}(0,\bm{x}) | \underline{\rm p}; \bm{x}_1, \bm{x}_2 \rangle^{(0)} 
\nonumber \\ 
  & \hspace{2cm}
    = \sum_{p \neq n} \delta^{(3)} (\bm{x}_1 - \bm{x}) \delta^{(3)} (\bm{x}_2 - \bm{x}) \; \langle \Omega | T^a \Phi_\ell^{\dag ab}(0,\bm{x}) | p; \bm{x}_1, \bm{x}_2 \rangle^{(0)}, 
\label{Wickoctet1}
\end{align} 
and 
\begin{align} 
& 
\sum_{k \neq n}  {}^{(0)}\langle \underline{\rm k}; \bm{x}'_1, \bm{x}'_2 | \Phi_\ell^{bc}(0,\bm{x}) \; \left( \psi^\dag T^c \chi \right) \!(\bm{x}) | \Omega \rangle 
\nonumber \\ 
  &\hspace{2cm}
    = \sum_{k \neq n} \delta^{(3)} (\bm{x}_1' - \bm{x}) \delta^{(3)} (\bm{x}_2' - \bm{x}) \; {}^{(0)}\langle  k; \bm{x}'_1, \bm{x}'_2  | \Phi_\ell^{bc}(0,\bm{x}) T^c | \Omega \rangle.  
\label{Wickoctet2}
\end{align} 
The ket state $T^a| p; \bm{x}_1, \bm{x}_2 \rangle^{(0)}$ stands for  $(T^a)_{ij}| p; \bm{x}_1, \bm{x}_2; j,i \rangle^{(0)}$
and the bra state ${}^{(0)}\langle  k; \bm{x}'_1, \bm{x}'_2  |$ $T^c$ for ${}^{(0)}\langle  k; \bm{x}'_1, \bm{x}'_2 ; i,j | (T^c)_{ij}$,
where the color indices have been assigned to the states according to footnote \ref{footindex}.
Note that necessarily both $|p; \bm{x}_1, \bm{x}_2 \rangle^{(0)}$  and  $|k; \bm{x}'_1, \bm{x}'_2  \rangle^{(0)}$ have to transform as octet states.
Plugging eqs.~\eqref{Wickoctet1} and~\eqref{Wickoctet2} into eq.~\eqref{eq:contact_hqoctet2}
and retaining only terms relevant for $P$-wave matrix elements, we obtain 
\begin{align} 
& \hspace{-3ex} 
- V_{{\cal O} ({}^1S_0^{[8]})} \delta^{(3)} (\bm{x}_1 - \bm{x}_1') \delta^{(3)} (\bm{x}_2 - \bm{x}_2') \Big|_{\textrm{$P$-wave}} 
\nonumber \\ 
&= 
\frac{1}{m^2} \delta^{(3)} (\bm{r}) \delta^{(3)} (\bm{x}_1 - \bm{x}_1') \delta^{(3)} (\bm{x}_2 - \bm{x}_2') 
\nonumber \\ & \hspace{3ex} \times 
\sum_{n \in {\mathbb S}} \sum_{p \neq n} \sum_{k \neq n} 
\langle \Omega | T^a \Phi_\ell^{\dag ab}(0,\bm{x}_1)  | p \rangle ^{(0)} 
\left( \frac{{}^{(0)} \langle p | g \bm{E}_1 | n \rangle^{(0)}}{(E_n^{(0)}-E_p^{(0)})^2} \cdot \ \overleftarrow{\bm{\nabla}}_1 
- \frac{{}^{(0)} \langle p | g \bm{E}_2^T | n \rangle^{(0)}}{(E_n^{(0)}-E_p^{(0)})^2} \cdot  \overleftarrow{\bm{\nabla}}_2 \right) 
\nonumber \\ 
& \hspace{9ex} \times 
\left(\bm{\nabla}_1\cdot \frac{{}^{(0)} \langle n | g \bm{E}_1 | k \rangle^{(0)}}{(E_n^{(0)}-E_k^{(0)})^2} 
- \bm{\nabla}_2\cdot \frac{{}^{(0)} \langle n | g \bm{E}_1 | k \rangle^{(0)}}{(E_n^{(0)}-E_k^{(0)})^2} \right) 
                {}^{(0)} \langle  k | \Phi_\ell^{bc}(0,\bm{x}_1) T^c | \Omega \rangle,
\label{eq:contact_hqoctet3} 
\end{align} 
where the derivatives $\bm{\nabla}_1$ and $\bm{\nabla}_2$ 
will eventually act on the wavefunctions
once eq.~\eqref{eq:contact_hqoctet3} is inserted into eq.~\eqref{eq:ldme_pnrqcd}.  
We suppress here and in the following the quark and antiquark positions in the eigenstates
when all the positions are the same: $\bm{x}_1' = \bm{x}_1$, $\bm{x}_2' = \bm{x}_2$, and $\bm{x}_1 = \bm{x}_2$.

The gluonic matrix elements of eq.~\eqref{eq:contact_hqoctet3} can be cast in the form of an integral over a temporal correlator of chromoelectric fields.
We proceed as follows. First, we write
\begin{align} 
  &\sum_{k \neq n} \frac{{}^{(0)} \langle n | g \bm{E}_1 | k \rangle^{(0)}}{(E_n^{(0)}-E_k^{(0)})^2} {}^{(0)} \langle  k| \Phi_\ell^{bc}(0,\bm{x}_1) T^c| \Omega \rangle 
\nonumber\\
  & \hspace{0.5cm}
    = -\sum_{k \neq n} \int_0^\infty dt \, t \,
    {}^{(0)}\langle n | \Phi_0(t,\bm{x}_1;0,\bm{x}_1) g \bm{E}_1(t)\Phi_0(0,\bm{x}_1;t,\bm{x}_1) | k \rangle^{(0)} \;
    {}^{(0)}\langle k | \Phi_\ell^{bc}(0,\bm{x}_1) T^c| \Omega \rangle
\nonumber \\ 
  & \hspace{0.5cm}
\underset{\textrm{for $\bm{r}=\bm{0}$}}{=} -\frac{1}{2 N_c} \int_0^\infty dt \, t \, {}^{(0)}\langle n| g \bm{E}_1^e(t) \, \Phi^{ec}_0 (0,\bm{x}_1;t,\bm{x}_1) \,  \Phi_\ell^{bc}(0,\bm{x}_1) 
    | \Omega \rangle. 
\label{eq:contact_hqoctet3aux} 
\end{align} 
In the first equality, we have replaced the ratio of the time independent matrix element to the square of the energy difference in left-hand side
with the time integral over the matrix element of the time dependent chromoelectric field supplemented, in a generic gauge, with the Schwinger line
\begin{equation}
\Phi_0 (t,\bm{x}_1;t',\bm{x}_1) = {\cal P} \exp\left[ -i g \int_t^{t'} d \tau \, A_0(\tau,\bm{x}_1) \right]\,,
\label{Schwinger}
\end{equation}
in the right-hand side.\footnote{
While eq.~\eqref{eq:contact_hqoctet3aux} is valid in any gauge, it can be
easily verified in the temporal gauge ($A_0 = 0$) where the Schwinger line
becomes unity. } 
Note that, whereas the Schwinger lines are path ordered in the color matrices, the fields in the matrix element are time ordered.
In the second line of eq.~\eqref{eq:contact_hqoctet3aux},
the Schwinger lines are in the fundamental representation, whereas, in the third one, the Schwinger line has to be understood in the adjoint representation.\footnote{
Consider an infinitesimal time interval $dt$ and fields located at a same but unspecified position, then it holds 
\begin{eqnarray*}
\Phi_0(t;t-dt) g \bm{E}(t)\Phi_0(t-dt;t) &=& \bm{E}^e(t)(T^e + ig\,dt\,A_0^b(t)[T^b,T^e]) = \bm{E}^e(t)(T^e + ig\,dt\,A_0^b(t)(if_{bed}T^d))
                                               \\
                                           &=& \bm{E}^e(t)T^d(\delta_{ed} -ig\,dt\,A_0^b(t)T^{b \, \rm adj}_{ed}) = \bm{E}^e(t)T^d  \Phi_0^{ed}(t-dt;t).
\end{eqnarray*}
See also ref.~\cite{Nayak:2005rt}.}
The condition $k\neq n$ in the first and second line of eq.~\eqref{eq:contact_hqoctet3aux} has been lifted for $\bm{r}=\bm{0}$ in the third line, since, according to footnote \ref{footindex2},
$n \in \mathbb S$ implies that $ {}^{(0)}\langle n | \Phi_\ell^{bc}(0,\bm{x}_1) T^c| \Omega \rangle$ vanishes after taking the color trace.
Because of this, the sum over the states $| k \rangle^{(0)}$ is complete and has been replaced with the identity in the last equality of eq.~\eqref{eq:contact_hqoctet3aux}.
Finally, in the last equality we have also computed the color trace:
${}^{(0)}\langle n;i,i|(\delta_{kj}/N_c)\,T^d_{kl}T^c_{lj} = {}^{(0)}\langle n;i,i| \delta^{dc}/(2N_c)$,
again using that ${}^{(0)}\langle  n |$ behaves like a color singlet for $\bm{r} = \bm{0}$.
Inserting eq.~\eqref{eq:contact_hqoctet3aux} and its Hermitian conjugate in eq.~\eqref{eq:contact_hqoctet3}, we arrive at 
\begin{align} 
\label{eq:contact_hqoctet4} 
& \hspace{-5ex} 
- V_{{\cal O} ({}^1S_0^{[8]})} \Big|_{\textrm{$P$-wave}} = - N_c \nabla_{\bm{r}}^i \delta^{(3)} (\bm{r}) \nabla_{\bm{r}}^j \frac{ {\cal E}^{ij} }{N_c^2 m^2} , 
\end{align} 
where the tensor ${\cal E}^{ij}$ is defined by 
\begin{align} 
\label{eq:ftensor} 
  {\cal E}^{ij} &=
                  \frac{1}{N_c}
                  \sum_n \int_0^\infty dt\, t \; \int_0^\infty dt'\, t' \;
                  \langle \Omega | \Phi_\ell^{\dag ab} (0,\bm{x}_1)
\,\Phi^{\dag ad}_0 (0,\bm{x}_1;t,\bm{x}_1) \, g {E}_1^{d,i}(t) | n \rangle^{(0)}\;
                  \nonumber\\
&\hspace{4cm}\times  
                  {}^{(0)} \langle n | g {E}_1^{e,j}(t') \,\Phi_0^{ec} (0,\bm{x}_1;t',\bm{x}_1) \,\Phi_\ell^{bc}(0,\bm{x}_1) | \Omega \rangle
\nonumber \\ 
&= 
\int_0^\infty dt\, t \; \int_0^\infty dt'\, t'\; \langle \Omega |
\Phi_\ell^{\dag ab} \Phi_0^{\dag ad} (0;t) g {E}^{d,i}(t) g {E}^{e,j}(t') \Phi_0^{ec} (0;t') \Phi_\ell^{bc} | \Omega \rangle.  
\end{align} 
Since $g {E}_1^{e,j}(t') \,\Phi_0^{ec} (0,\bm{x}_1;t',\bm{x}_1) \,\Phi_\ell^{bc}(0,\bm{x}_1)$ does not contain color matrices,
the matrix element ${}^{(0)} \langle n | g {E}_1^{e,j}(t') \,\Phi_0^{ec} (0,\bm{x}_1;t',\bm{x}_1) \,\Phi_\ell^{bc}(0,\bm{x}_1) | \Omega \rangle$ vanishes for $n \notin {\mathbb S}$.  
Hence, we could extend the sum in the first line of eq.~\eqref{eq:ftensor} to all $n$ and use the completeness of the eigenstates $| n \rangle^{(0)}$ to replace their sum with the identity
operator in the second equality: $\displaystyle \sum_n |n;i,i\rangle^{(0)}\,{}^{(0)}\langle n; j,j| = \delta_{ij}\delta_{ij} = N_c$.
Because of translational invariance, we have dropped in the last line of eq.~\eqref{eq:ftensor} the space coordinate, which is the same, but arbitrary, for all operators there.

\begin{figure}[ht]
\begin{center}
\includegraphics[width=0.45\textwidth]{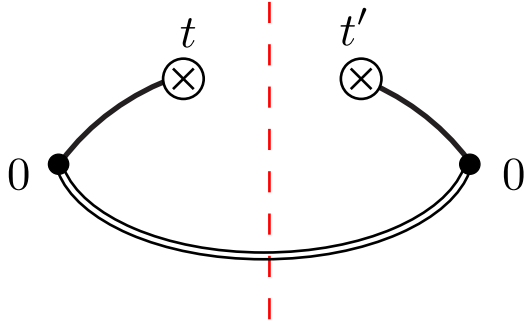} 
\caption{\label{fig:wilsonlines}
Graphical representation of the chromoelectric fields and Wilson lines in the integrand of eq.~\eqref{eq:ftensor}. 
The symbols $\otimes$ represent insertions of chromoelectric fields at the times $t$ and $t'$, and the filled circles represent the spacetime origin.
Solid lines are Schwinger lines, double lines are gauge-completion Wilson 
lines in the $\ell$ direction, and the dashed line is the cut.
}
\end{center}
\end{figure}

In eq.~\eqref{eq:ftensor}, the chromoelectric field at time $t'$ is connected to the origin $0$ by the Schwinger line $\Phi_0^{ec} (0;t')$,
which then continues to infinity in the $\ell$ direction.
Analogously, the chromoelectric field at time $t$ is connected to the origin $0$ by the Schwinger line $\Phi_0^{\dag ad} (0;t)$, which then continues to infinity in the $\ell$ direction.  
For a suitable choice of the sign of $\ell^0$, the fields in $g {E}^{e,j}(t') \Phi_0^{ec} (0;t') \Phi_\ell^{bc}$ are time ordered (${\cal T}$),
and those in $\Phi_\ell^{\dag ab} \Phi_0^{\dag ad} (0;t) g {E}^{d,i}(t)$ are anti-time ordered ($\bar{\cal T}$).
Hence, eq.~\eqref{eq:ftensor} can be interpreted as a cut diagram, which can be useful for perturbative QCD.  
We show this configuration of Wilson lines graphically in figure~\ref{fig:wilsonlines}. 

We derive now two properties of the (time ordered) operator $g {E}^{e,i}(t) \Phi_0^{ec} (0;t) \Phi_\ell^{bc}$.
First, the Hermitian conjugate of $g {E}^{e,i}(t) \Phi_0^{ec} (0;t) \Phi_\ell^{bc}$ is 
\begin{align}
(g {E}^{e,i}(t) \Phi_0^{ec} (0;t) \Phi_\ell^{bc})^\dag &= ({\cal T}\, g {E}^{e,i}(t) \Phi_0^{ec} (0;t) \Phi_\ell^{bc})^\dag
                                                          = \bar{\cal T} \Phi_\ell^{\dag cb} \Phi_0^{\dag ce} (0;t) g {E}^{e,i}(t)
\nonumber\\
                                                          &= \Phi_\ell^{\dag cb} \Phi_0^{\dag ce} (0;t) g {E}^{e,i}(t),
\end{align}
where we have dropped the time ordering prescription whenever the operators appear in the right time or anti-time ordering. 
This guarantees that the operator $\Phi_\ell^{\dag ab} \Phi_0^{\dag ad} (0;t) g {E}^{d,i}(t)$ $g {E}^{e,i}(t) \Phi_0^{ec} (0;t) \Phi_\ell^{bc}$ is Hermitian.
Second, under a gauge transformation $U(\theta(x))$, the operator  $g {E}^{e,i}(t') \Phi_0^{ec} (0;t') \Phi_\ell^{bc}$
transforms as an adjoint field strength tensor at infinity in the $\ell$ direction:
\begin{align}
g {E}^{e,i}(t) \Phi_0^{ec} (0;t) \Phi_\ell^{bc} \xrightarrow[U(\theta(x))]{}
& {\cal T}  g {E}^{e',i}(t)U^{\dag e'e}(\theta(t)) \, U^{ee''}(\theta(t))\Phi_0^{e''c''} (0;t) U^{\dag c''c}(\theta(0)) \, 
                                                  \nonumber\\
& \hspace{3.8cm} \times U^{bb'}(\theta(\ell\infty)) \Phi_\ell^{b'c'} U^{\dag c'c}(\theta(0))   
                                                   \nonumber\\
& = {\cal T} g {E}^{e',i}(t) \Phi_0^{e'c''} (0;t)  U^{\dag c''c}(\theta(0))
U^{\dag c'c}(\theta(0))  \Phi_\ell^{b'c'} U^{bb'}(\theta(\ell\infty)) 
                                                   \nonumber\\
& = g {E}^{e',i}(t) \Phi_0^{e'c'} (0;t)   \Phi_\ell^{b'c'} U^{bb'}(\theta(\ell\infty)).   
\end{align}
After the first equality we have explicitly required that the fields are time ordered, and in the last equality we have used that $U^{\dag} = U^{-1}$,
because $U$ is unitary, and $U^{\dag}=U^T$, because $U$ is real in the adjoint representation.\footnote{
The SU(3) generators in the adjoint representation are purely imaginary.} 
The gauge transformation property of the operator $g {E}^{e,i}(t') \Phi_0^{ec} (0;t') \Phi_\ell^{bc}$ guarantees the gauge invariance of the tensor ${\cal E}^{ij}$.

Having matched the color-octet contact term, we can compute the color-octet matrix element $\langle \Omega  |{\cal O}^{h_Q} ({}^1S_0^{[8]}) | \Omega \rangle$ from eq.~\eqref{eq:ldme_pnrqcd}.  
Since the product $\phi^{(0)}_{^1P_1} (\bm{r}) \phi^{(0)\,*}_{^1P_1}{} (\bm{r})$  is isotropic after summing over the polarizations of the $^1P_1$ state, 
the tensor $\nabla_{\bm{r}}^i \nabla_{\bm{r}}^j$ in eq.~\eqref{eq:contact_hqoctet4} can be replaced by $\bm{\nabla}_{\bm{r}} \cdot \bm{\nabla}_{\bm{r}} \delta^{ij}/3$.  
Then, we obtain at leading nonvanishing order in $v$
\begin{equation} 
\label{eq:hqoctet} 
\langle \Omega  |{\cal O}^{h_Q} ({}^1S_0^{[8]}) | \Omega \rangle = 3 \times \frac{3 N_c}{2 \pi} |R^{(0)}{}'(0)|^2 \frac{{\cal E}}{9 N_c m^2}, 
\end{equation} 
where ${\cal E}$ is the dimensionless gluonic correlator 
\begin{equation} 
\label{eq:ecorrelator} 
{\cal E} = \frac{3}{N_c} \int_0^\infty dt\, t \; \int_0^\infty dt'\, t' \;
\langle \Omega | \Phi_\ell^{\dag ab} \Phi_0^{\dag ad} (0;t) g {E}^{d,i}(t) g {E}^{e,i}(t') \Phi_0^{ec} (0;t') \Phi_\ell^{bc} | \Omega \rangle.  
\end{equation} 
The correlator ${\cal E}$ is the isotropic part of ${\cal E}^{ij}$, i.e. $(N_c\delta^{ij}/9)\, {\cal E}$.  
The factor $3/N_c$ in the definition of ${\cal E}$ has been chosen so that eq.~\eqref{eq:hqoctet} resembles the pNRQCD expression
for the color-octet decay matrix element~\cite{Brambilla:2001xy, Brambilla:2002nu} 
\begin{equation} 
\label{eq:hqoctet_decay} 
\langle h_Q | \psi^\dag T^a \chi \chi^\dag T^a \psi | h_Q \rangle = \frac{3 N_c}{2 \pi} |R^{(0)}{}'(0)|^2 \frac{{\cal E}_3}{9 N_c m^2}, 
\end{equation} 
where the correlator ${\cal E}_3$ is defined by 
\begin{equation} 
\label{eq:E3def} 
{\cal E}_3 = \frac{1}{2 N_c} \int_0^\infty dt\, t^3 \, \langle \Omega | g {E}^{a,i}(t) \Phi_0^{ab} (0;t) g {E}^{b,i}(0) | \Omega \rangle.  
\end{equation} 

The contact terms for the operators involving the $\chi_{QJ}$ can be computed in a similar way.
We obtain 
\begin{align} 
\label{eq:contact_chi} 
- V_{{\cal O} ({}^3P_J^{[1]})} \Big|_{\textrm{$P$-wave}} 
&=  - T_{1J}^{ij} N_c \nabla_{\bm{r}}^i \delta^{(3)} (\bm{r}) \nabla_{\bm{r}}^j, 
\\ 
\label{eq:contact_3S1} 
- V_{{\cal O} ({}^3S_1^{[8]})} \Big|_{\textrm{$P$-wave}} 
&= - \sigma^k \otimes \sigma^k N_c \nabla_{\bm{r}}^i \delta^{(3)} (\bm{r}) \nabla_{\bm{r}}^j \frac{{\cal E}^{ij}}{N_c^2 m^2}, 
\end{align} 
where we have again displayed only the terms relevant for the $P$-wave LDMEs.
The tensors $T_{1J}^{ij}$ are spin projectors that are defined by 
\begin{subequations} 
\begin{align} 
T_{10}^{ij} &= \frac{1}{3} \sigma^i \otimes \sigma^j, 
\\ 
T_{11}^{ij} &= \frac{1}{2} \epsilon_{kim} \epsilon_{kjn} \sigma^m \otimes \sigma^n, 
\\ 
T_{12}^{ij} &= \left( \frac{\delta_{im} \sigma^n + \delta_{in} \sigma^m}{2} - \frac{\delta_{mn}}{3} \sigma^i \right) \otimes \left( \frac{\delta_{jm} \sigma^n + \delta_{jn} \sigma^m}{2} 
- \frac{\delta_{mn}}{3} \sigma^j \right).  
\end{align} 
\end{subequations}

The wavefunction of a $^3P_J$ state with polarization $\lambda$ is given by 
\begin{equation} 
  \phi^{(0)}_{^3P_J} (\bm{r}) = \sum_{M,S} R^{(0)} (r) Y_1^M (\bm{\hat r}) |1 S\rangle \langle 1 M ; 1 S | J \lambda \rangle , 
\end{equation} 
where $|1 S\rangle$ is a spin-triplet state, and $\langle 1 M ; 1 S | J \lambda \rangle$ are Clebsch--Gordan coefficients.  
At leading order in $v$, the radial wavefunction $R^{(0)}(r)$ does not depend on the heavy quark spin.  

Inserting the above formulas in eq.~\eqref{eq:ldme_pnrqcd}, we obtain the following expressions for the LDMEs, which are valid at leading nonvanishing order in $v$,
\begin{subequations} 
\label{eq:ldmes_chi} 
\begin{align} 
\langle \Omega | {\cal O}^{\chi_{QJ}}({}^3P_J^{[1]}) | \Omega \rangle 
&= (2 J+1) \times \frac{3 N_c}{2\pi} |R^{(0)}{}'(0)|^2, 
\label{eq:ldmes_chi_cs} 
\\ 
\langle \Omega | {\cal O}^{\chi_{QJ}}({}^3S_1^{[8]}) | \Omega \rangle 
&= (2 J+1) \times \frac{3 N_c}{2 \pi} |R^{(0)}{}'(0)|^2 \frac{{\cal E}}{9 N_c m^2}.   
\label{eq:ldmes_chi_co} 
\end{align} 
\end{subequations}
Together with eqs.~\eqref{eq:hqsinglet} and \eqref{eq:hqoctet}, they enter the NRQCD factorization formulas for production of $P$-wave quarkonia at leading order in $v$.
Note that $\langle \Omega | {\cal O}^{\chi_{QJ}}({}^3S_1^{[8]}) | \Omega \rangle$ depends on $\cal E$ and not on ${\cal E}^{ij}$ because the matrix element is polarization summed.
The results for the color-singlet and color-octet matrix elements lead to the relations
\begin{align} 
\label{eq:ratio} 
  \frac{m^2\,\langle \Omega | {\cal O}^{\chi_{QJ}}({}^3S_1^{[8]}) | \Omega \rangle} {\langle \Omega | {\cal O}^{\chi_{QJ}}({}^3P_J^{[1]}) | \Omega \rangle} 
= \frac{m^2\,\langle \Omega | {\cal O}^{h_Q}({}^1S_0^{[8]}) | \Omega \rangle}{\langle \Omega | {\cal O}^{h_Q}({}^1P_1^{[1]}) | \Omega \rangle} 
= \frac{{\cal E}}{9 N_c}.
\end{align} 
These relations are valid for any $P$-wave quarkonium state, since the right-hand side is independent of the flavor and of the principal quantum number.\footnote{
That the ratio $\langle \Omega | {\cal O}^{\chi_{QJ}}({}^3S_1^{[8]}) | \Omega \rangle/\langle \Omega | {\cal O}^{\chi_{QJ}}({}^3P_J^{[1]}) | \Omega \rangle$
    scales like $1/m^2$ times a constant that is independent of the flavor and the principal quantum number has been assumed on phenomenological grounds in ref.~\cite{Likhoded:2012hw}.
}  
It follows that all NRQCD matrix elements for production of $P$-wave quarkonia at leading order in $v$ are fixed once the quantity ${\cal E}$
and the wavefunctions at the origin are known.
By using eqs.~\eqref{eq:hqsinglet}, \eqref{eq:hqoctet}, \eqref{eq:ldmes_chi_cs}, and \eqref{eq:ldmes_chi_co},
we obtain the following pNRQCD expressions for the production rates of $P$-wave quarkonia 
\begin{subequations} 
\label{eq:fac_pwave_pNRQCD} 
\begin{align} 
\sigma_{h_Q+X} &= 
3 \left( \sigma_{Q \bar Q({}^1P_1^{[1]})} + \sigma_{Q \bar Q({}^1S_0^{[8]})} \frac{{\cal E}}{9 N_c m^2} \right) \frac{3 N_c}{2 \pi} | R^{(0)}{}'(0)|^2, 
\label{eq:fac_pwave_pNRQCDa}
  \\ 
\sigma_{\chi_{QJ}+X} &= 
(2 J+1) \left( \sigma_{Q \bar Q({}^3P_J^{[1]})} + \sigma_{Q \bar Q({}^3S_1^{[8]})} \frac{{\cal E}}{9 N_c m^2} \right) \frac{3 N_c}{2 \pi} | R^{(0)}{}'(0)|^2, 
\label{eq:fac_pwave_pNRQCDb}
\end{align} 
\end{subequations} 
which are valid at leading order in $v$ and up to corrections of order $1/N_c^2$.

\subsection{Consistency with NRQCD factorization} 
\label{sec:consistency}
As we have discussed earlier, the validity of NRQCD factorization requires the short-distance coefficients to be free of infrared divergences.  
The short-distance coefficients are determined by replacing the heavy quarkonium state ${\cal Q}$ in the NRQCD factorization formula \eqref{eq:NRQCDfac}
by a perturbative $Q \bar Q$ state, and computing both sides as perturbation series in $\alpha_{\text{s}}$.  
Hence, for the NRQCD factorization to be valid, the infrared divergences on the left-hand side of eq.~\eqref{eq:NRQCDfac} must be reproduced exactly by the LDMEs on the right-hand side.  
For our expressions of the LDMEs and perturbative QCD calculations of the short-distance coefficients to be consistent,
the pNRQCD expressions of the LDMEs, when computed in perturbative QCD,
must reproduce the infrared divergences that appear in the LDMEs when the quarkonium states are replaced by perturbative $Q \bar Q$ states.
This is nontrivial in the case of the color-octet LDMEs, which involve arbitrary exchanges of gluons between the quark, antiquark, and the gauge-completion Wilson lines.

\begin{figure}[ht]
\begin{center}
\includegraphics[width=0.8\textwidth]{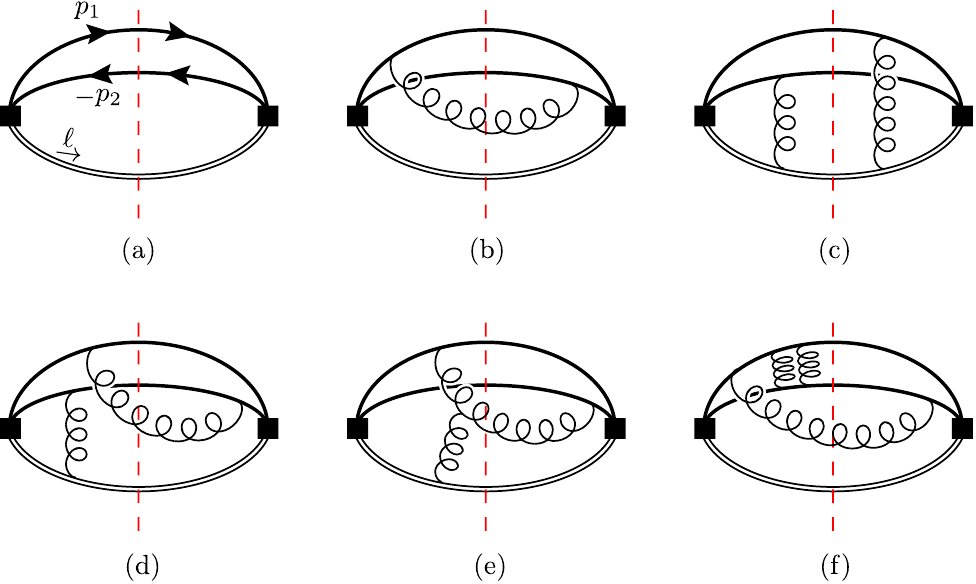}
\caption{\label{fig:feynman_diagrams}
  Representative Feynman diagrams for the color-octet LDME $\langle \Omega | {\cal O}^{Q(p_1) \bar{Q}(p_2)} ({}^3S_1^{[8]})| \Omega \rangle$ computed in perturbative QCD at
  (a) LO, (b) NLO, and (c)--(e) NNLO in $\alpha_{\text{s}}$.
  Solid lines are quark lines, double lines are the gauge-completion Wilson lines in the adjoint representation in the $\ell$ direction, curly lines are gluons,
  dashed lines are final-state cuts, and filled squares represent operators which create a $Q \bar Q$ pair in a color-octet state.
  Diagram (f) shows a NNNLO contribution that contains a logarithmic IR divergence, which can be absorbed into the quarkonium wavefunctions.
}
\end{center}
\end{figure}

Well-established methods to investigate infrared divergences in perturbative QCD calculations have been developed in proofs of perturbative
factorization~\cite{Collins:1981ta, Collins:1989gx, Collins:1985ue,Collins:1988ig, Bodwin:1984hc}.
This consists of applying approximations to gluon propagators and vertices that simplify the perturbative QCD expressions while preserving the infrared divergences,
and reorganizing the resulting expressions in terms of Wilson lines.
In ref.~\cite{Nayak:2005rt}, the authors considered the infrared divergences in the color-octet LDME $\langle \Omega | {\cal O}^{Q(p_1) \bar{Q}(p_2)} ({}^3S_1^{[8]})| \Omega \rangle$. 
In figure~\ref{fig:feynman_diagrams}, we show some representative Feynman diagrams that appear in the perturbative QCD calculation of the LDME. 
The quark and antiquark in the final state are on shell and have momenta $p_1 = p+q$ and $p_2 = p-q$, respectively.
There are additional diagrams that can be obtained by moving gluon attachments from one side of the cut to the other side,
or by replacing gluon attachments to the quark (antiquark) line by attachments to the antiquark (quark) line.
Note that, since the scale $m$ has been integrated out in NRQCD, gluons cannot have momenta that exceed scales of order $mv$.
In these diagrams, infrared divergences arise when the virtual gluons have soft momenta, that is, when the temporal and spatial components of the gluon momenta are small and of the same order.
In such case, the soft approximation can be used for propagators and vertices involving these gluons~\cite{Grammer:1973db, Collins:1981uk, Collins:1989gx}.

When the quark and antiquark in the final state are on shell, the soft approximation leads to the eikonal approximation,
where a soft gluon with momentum $k$ has propagator and vertex given in Feynman gauge by $i/(\beta \cdot k + i \varepsilon)$ and $\mp i g_s T^a \beta^\mu$,
where $\beta$ is the four-velocity of the quark or antiquark line, and the minus (plus) sign applies to the quark (antiquark) line.
Note that the vertex factor in the eikonal approximation does not involve gamma matrices.
It is clear that the Feynman rules for eikonal gluon attachments to the quark line are equivalent to the Feynman rules
for gluon attachments to the path-ordered Wilson line in the fundamental representation in the direction of the quark momentum.
Similarly, eikonal gluon attachments to the antiquark line are equivalent to gluon attachments to anti-path-ordered Wilson line in the direction of the antiquark momentum.
That is, the eikonal approximation decouples soft gluon attachments to quark and antiquark lines, and factors them out in the form of Wilson lines.
If the $Q \bar Q$ in the final state is in a color-singlet state,
we obtain the following infrared factor by collecting all Wilson-line factors arising from the application of the eikonal approximation 
to soft gluon attachments to the quark and antiquark lines on both sides of the cut:
\begin{equation}
\label{eq:inffactor_allorder}
{\cal I} (p, q) = \sum_N \langle \Omega |{\cal \bar {T}} \Big\{ {\rm Tr} [ \bar{\Phi}_{p_2}^{\dag} T^a\Phi_{p_1}^\dag ] \Phi_\ell^\dag{}^{ab} \Big\}| N \rangle
\langle N |{\cal T} \Big\{\Phi_\ell^{bc} {\rm Tr} [ \Phi_{p_1} T^c \bar{\Phi}_{p_2}]\Big\}| \Omega \rangle,
\end{equation}
where $\displaystyle \Phi_{p_1} = {\cal P} {\rm exp} \left[ -i g_s \int_0^\infty d \lambda \, p_1 \cdot A(p_1 \lambda)\right]$ is the path-ordered Wilson line in the fundamental representation
in the $p_1$ direction, and $\displaystyle \bar{\Phi}_{p_2}= \bar{{\cal P}}  {\rm exp} \left[ i g_s \int_0^\infty d \lambda \, p_2 \cdot A(p_2 \lambda)\right]$
is the anti-path-ordered Wilson line in the fundamental representation in the $p_2$ direction.
The sum over $N$ contains all possible intermediate states, and the traces are over color indices in the fundamental representation. 
All infrared divergences associated with soft gluons that appear in the color-octet LDME is factored out into the infrared factor ${\cal I} (p_1, p_2)$,
so that $\langle \Omega | {\cal O}^{Q(p_1) \bar{Q}(p_2)} ({}^3S_1^{[8]})| \Omega \rangle / {\cal I} (p_1, p_2)$ is free of such divergences.
This infrared factor has first been obtained in Refs.~\cite{Nayak:2005rw,Nayak:2005rt} in the analysis of gluon fragmentation functions for the process $g \to Q \bar Q + X$.

The quantity $\langle \Omega | {\cal O}^{Q(p_1) \bar{Q}(p_2)} ({}^3S_1^{[8]})| \Omega \rangle / {\cal I} (p_1, p_2)$ is {\it not} completely free of infrared divergences,
because the LDME can have additional infrared divergences from gluon exchanges between the quark and antiquark lines.
These infrared divergences are not captured by the infrared factor~(\ref{eq:inffactor_allorder}),
because they arise from the region of gluon momentum where its temporal component is much smaller than the spatial components;
that is, they are singularities associated with potential gluons.
From proofs of perturbative factorization in quarkonium decays in NRQCD, it can be seen that at leading order in $v$  
infrared divergences arising from gluon exchanges between the quark and antiquark in a color-singlet state can be absorbed into the quarkonium wavefunctions at the origin~\cite{Bodwin:1994jh}.
Logarithmic divergences arising from quarkonium wavefunctions at the origin appear from two loops, and hence, these divergences affect the color-octet LDME from three loops,
because at least one extra gluon is needed in order to carry the color charge over the cut when the $Q \bar Q$ is in a color-singlet state.
The Feynman diagram in figure~\ref{fig:feynman_diagrams}(f) shows a contribution at NNNLO in $\alpha_{\text{s}}$ from which such logarithmic divergences appear.
It is possible that there are additional infrared divergences that cannot be absorbed into the quarkonium wavefunctions at the origin nor the infrared factor,
if there are infrared divergences arising from soft gluons interacting with potential gluons.
Nonetheless, such effects do not appear at least until three-loop accuracy, as can be seen in explicit calculations of the color-octet LDME in perturbative QCD~\cite{Nayak:2006fm,Bodwin:2019bpf}.
Therefore, the infrared factor does contain all infrared divergences appearing in the color-octet LDME to two-loop accuracy, which come from soft gluons carrying momenta of order $mv$.

It is particularly interesting to investigate the infrared factor~(\ref{eq:inffactor_allorder}) at the lowest nonvanishing power in the relative momentum $q$,
because matching calculations are usually carried out at a specific accuracy in $q/m$.
The lowest power term is of quadratic order in $q$, where a factor of $q$ comes from each side of the cut.
To relative order $\bm{q}^2/m^2$, the infrared factor is given by ${\cal I}_2 (p,q)$, which is defined in eq.~(32) of ref.~\cite{Nayak:2005rt} as\footnote{
We correct a typo by replacing  $\Phi_\ell^{\dag bc'}$ with  $\Phi_\ell^{\dag c'b}$.
Moreover, we make explicit the time and anti-time orderings, which are implicit in ref.~\cite{Nayak:2005rt}.}
\begin{align} 
\label{eq:irfac} 
{\cal I}_2 (p,q) \equiv & 
\sum_N \int_0^\infty d \lambda' \, \lambda' 
\langle \Omega | \bar{\cal T} \left\{ \Phi_\ell^{\dag c'b} \Phi_p^{\dag a'c'}(\lambda') [p^\mu q^\nu G_{\nu \mu}^{a'} (\lambda'p) ] \right\} | N \rangle 
\nonumber \\ 
& \times 
\int_0^\infty d \lambda \, \lambda \langle N | {\cal T} \left\{ \Phi_\ell^{bc} [ p^\mu q^\nu G_{\nu \mu}^{a} (\lambda p)] \Phi_p^{ac} (\lambda) \right\} | \Omega\rangle,
\end{align} 
where 
\begin{align} 
\Phi_p (\lambda) = {\cal P} \exp \left[ - i g \int_0^\lambda d \lambda' \, p \cdot A^{\rm adj} (p\lambda') \right], 
\end{align} 
is an adjoint Wilson line along $p$.
Since in eq.~\eqref{eq:irfac} a momentum $q$ comes from each side of the cut in the squared amplitude, the infrared factor contributes to the production of a color-singlet $P$-wave state.  
The same infrared factor appears in the fragmentation of a gluon into color-singlet $Q \bar Q$. 
Since the gluon fragmentation process produces a $Q \bar Q$ in a color-octet $^3S_1$ state,
the divergences must match the infrared divergences in the color-octet LDME $\langle \Omega | {\cal O}^{\cal Q} (^3S_1^{[8]}) | \Omega \rangle$,
when $\cal Q$ is replaced by a color-singlet $Q \bar Q$ state.  
This agreement has been confirmed explicitly through one-loop and partial two-loop calculations of the color-octet LDME in ref.~\cite{Nayak:2006fm};
the two-loop calculations have only been done for the diagrams involving the gauge-completion Wilson lines.  
Since the LDME $\langle \Omega | {\cal O}^{\cal Q}(^3S_1^{[8]}) | \Omega \rangle$ appears in the NRQCD factorization formula at leading order in $v$,
the same infrared divergences have to occur in the operator $\langle \Omega | {\cal O}^{\cal Q} (^1S_0^{[8]}) | \Omega \rangle$ because of the heavy quark spin symmetry.  

In the rest frame of the $Q \bar Q$, where $\bm{p} = 0$ and $q^0 = 0$,
$\Phi_p (\lambda)$ is the Schwinger line $\Phi_0 (0,\bm{0}; t,\bm{0})$ in the SU(3) adjoint representation with $t = \sqrt{p^2} \lambda$,
and $p^\mu q^\nu G_{\nu \mu}^{a} (\lambda p)$  is $-\sqrt{p^2} q^i E^{a \, i} (t)$ with the chromoelectric field located at $\bm{0}$.  
Therefore, the infrared factor ${\cal I}_2(p,q)$ defined in eq.~\eqref{eq:irfac} can be written in the rest frame of the $Q \bar Q$ as
\begin{align} 
\label{eq:irfac2} 
{\cal E}^{ij} \frac{q^i q^j}{p^2}, 
\end{align} 
where ${\cal E}^{ij}$ is the tensor defined in eq.~\eqref{eq:ftensor}.  
Expression~\eqref{eq:irfac2} is proportional to the contact terms $V_{{\cal O}(^1S_0^{[8]})}$ and $V_{{\cal O}(^3S_1^{[8]})}$ in eqs.~\eqref{eq:contact_hqoctet4} and \eqref{eq:contact_3S1}, 
respectively, when written in momentum space for a color-singlet $Q \bar Q$ state with relative momentum $\bm{q}$.  
The pNRQCD expressions for the color-octet LDMEs in eqs.~\eqref{eq:hqoctet} and \eqref{eq:ldmes_chi_co} are, therefore,
expected to reproduce the same infrared divergences obtained from the NRQCD factorization.  

This is straightforward to check at one-loop accuracy.  
By computing the correlator ${\cal E}$ at order $\alpha_{\text{s}}$ accuracy in dimensional regularization with $d=4-2 \epsilon$ spacetime dimensions, we obtain 
\begin{align} 
\label{eq:corrE} 
{\cal E} = 6 C_F \frac{\alpha_{\text{s}}}{\pi} \left( \frac{1}{\epsilon_{\rm UV}} - \frac{1}{\epsilon_{\rm IR}} \right) +O(\alpha_{\text{s}}^2), 
\end{align} 
where $C_F= (N_c^2-1)/(2N_c)$ is the Casimir of the fundamental representation of SU(3)
and the subscripts UV and IR indicate the origin (ultraviolet and infrared, respectively) of the $1/\epsilon$ poles.  
After renormalizing the UV divergence at the scale $\Lambda$, ${\cal E}$ satisfies the following renormalization group equation 
\begin{equation} 
\label{eq:corrE_RG} 
\frac{d}{d \log \Lambda} {\cal E} (\Lambda) = 12 C_F \frac{\alpha_{\text{s}}}{\pi} + O(\alpha_{\text{s}}^2),
\end{equation} 
which, in turn, implies the following renormalization group equation for the NRQCD matrix elements (see eqs.~\eqref{eq:ldmes_chi_cs} and~\eqref{eq:ldmes_chi_co})
\begin{equation} 
\label{eq:RG} 
\frac{d}{d \log \Lambda} \langle {\cal O}^{\chi_{QJ}} ({}^3S_1^{[8]}) \rangle = \frac{4 C_F \alpha_{\text{s}}}{3 N_c \pi m^2} \langle {\cal O}^{\chi_{QJ}} ({}^3P_J^{[1]}) \rangle.  
\end{equation} 
The same evolution equation relates $\langle {\cal O}^{h_{Q}} ({}^1S_0^{[8]}) \rangle$ with $\langle {\cal O}^{h_{Q}} ({}^1P_1^{[1]}) \rangle$.  
Equation~\eqref{eq:RG} agrees with the evolution equation derived from a perturbative calculation in NRQCD~\cite{Bodwin:1994jh},
and, therefore, the UV divergence at one-loop accuracy of the color-octet LDME in pNRQCD is the same as in NRQCD.  
Since loop corrections to NRQCD matrix elements are scaleless, UV poles cancel IR poles in eq.~\eqref{eq:corrE}.
Hence, the IR divergence at one-loop accuracy of the color-octet LDME in pNRQCD is the same as in NRQCD too.

At two loops, some consistency checks between the pNRQCD result and the NRQCD factorization can be made based on the two-loop calculations in refs.~\cite{Nayak:2005rt} and~\cite{Nayak:2006fm}. 
In ref.~\cite{Nayak:2005rt}, two-loop corrections to the infrared factor ${\cal I}_2(p,q)$ associated with the gauge-completion Wilson lines were computed. 
These contribute to the infrared divergences of the LDMEs  $\langle {\cal O}^{h_{Q}} ({}^1S_0^{[8]}) \rangle$ and $\langle {\cal O}^{\chi_{QJ}} ({}^3S_1^{[8]}) \rangle$ at order $\alpha_{\text{s}}^2$,
and were reproduced in ref.~\cite{Nayak:2006fm} through the explicit calculation of the LDMEs.  
Since, as we have seen, the calculation of the infrared factor ${\cal I}_2(p,q)$ is equivalent to the calculation of the infrared divergences
in the contact terms $V_{{\cal O}(^1S_0^{[8]})}$ and $V_{{\cal O}(^3S_1^{[8]})}$, and eventually in the chromoelectric field tensor ${\cal E}^{ij}$,
it follows that the pNRQCD expressions for the color-octet LDMEs have the same infrared divergences associated with the gauge-completion Wilson lines
as those found in the NRQCD calculations of ref.~\cite{Nayak:2006fm}.  

It may be interesting to note that at order $\alpha_{\text{s}}$ the pole structure of the correlator ${\cal E}$ and of the correlator ${\cal E}_3$,
defined in eq.~\eqref{eq:E3def} and relevant for the decay widths of $P$-wave quarkonia, is the same, i.e. the one given in eq.~\eqref{eq:corrE}.
This reflects at the pNRQCD level the fact that at the NRQCD level the one-loop renormalization group equation~\eqref{eq:RG} is the same as the one-loop renormalization group equation
for the decay matrix elements appearing in inclusive decays of $P$-wave quarkonia~\cite{Bodwin:1994jh}.  
The observation, however, ceases to hold at two loops, because at this order, ${\cal E}$ receives contributions from the gauge-completion Wilson lines, which are absent in ${\cal E}_3$.\footnote{
The difference between ${\cal E}$ and ${\cal E}_3$ signals also the violation of the crossing symmetry beyond leading order in the corresponding NRQCD matrix elements~\cite{Bodwin:1994jh}.} 

An important issue in NRQCD factorization is whether the color-octet LDMEs are independent of the direction of the gauge-completion Wilson lines,
which is necessary in establishing the universality of the NRQCD production matrix elements.  
While a general argument for the universality has been suggested in ref.~\cite{Nayak:2015qca},
an explicit verification has only been done at two-loop
accuracy~\cite{Nayak:2005rt,Nayak:2006fm, Bodwin:2019bpf, Zhang:2020atv}.  
In the pNRQCD expression of the color-octet LDMEs, the dependence on the direction of the gauge-completion Wilson lines is encoded in the tensor ${\cal E}^{ij}$.
For the case of polarization-summed cross sections, where the polarization of the quarkonium in the final state is summed over, only the isotropic part of ${\cal E}^{ij}$, 
given by $(N_c\delta^{ij}/9)\, {\cal E}$, contributes to the color-octet LDMEs, and, therefore,
the dependence on the direction of the gauge-completion Wilson lines disappears due to rotational symmetry.  
Hence, the pNRQCD expressions of the color-octet LDMEs support the universality of the NRQCD LDMEs for polarization-summed cross sections of $P$-wave quarkonia.

\section{\boldmath Phenomenology  of inclusive production of $P$-wave quarkonia} 
\label{sec:pwave-phen}

\subsection[Phenomenological determination of $\cal E$]{\boldmath Phenomenological determination of $\cal E$} 
\label{sec:corrE_determination} 
In order to compute the cross sections of $P$-wave quarkonia,
it is necessary to obtain values for the radial wavefunctions at the origin and the gluonic correlator ${\cal E}$.
Radial wavefunctions at the origin can be computed by solving the Schr\"odinger equation or can be determined from decay rates.
The gluonic correlator ${\cal E}$ could, in principle, be computed from lattice QCD.
Since, however, a lattice QCD determination of ${\cal E}$ is not available at the time of this study, 
we determine ${\cal E}$ phenomenologically by comparing the measured ratio of $P$-wave quarkonium cross sections to the expression obtained from the NRQCD/pNRQCD factorization.

We use the ratio $r_{21}$ of the $\chi_{c2}(1P)$ and $\chi_{c1}(1P)$ differential cross sections defined by 
\begin{equation} 
r_{21} = \frac{d \sigma_{\chi_{c2}(1P)}/dp_T} {d \sigma_{\chi_{c1}(1P)}/dp_T}, 
\end{equation} 
where $p_T$ is the transverse momentum of the $\chi_{cJ}(1P)$.  
This ratio has been measured at the LHC by CMS~\cite{Chatrchyan:2012ub} and ATLAS~\cite{ATLAS:2014ala}.  
In order to compute the cross sections $\sigma_{\chi_{cJ}(1P)}$ in NRQCD, we employ the short-distance coefficients $\sigma_{Q \bar Q(^3P_J^{[1]})}$ and $\sigma_{Q \bar Q(^3S_1^{[8]})}$
that were computed in ref.~\cite{Bodwin:2015iua} at next-to-leading order (NLO) in $\alpha_{\text{s}}$ for a center-of-mass energy of 7~TeV with rapidity range $|y|<0.75$.
The calculation in ref.~\cite{Bodwin:2015iua} used for the charm quark mass $m_c = 1.5$~GeV, CTEQ6M parton distribution functions at the scale $\mu_F = \sqrt{p_T^2 + 4 m_c^2}$, 
and computed $\alpha_{\text{s}}$ at the same scale $\mu_R = \sqrt{p_T^2 + 4 m_c^2}$,
running at two loops with $n_f = 5$ light quark flavors and $\Lambda_{\rm QCD}^{(5)} = 226$~MeV.  
The short-distance coefficients $\sigma_{Q \bar Q(^3P_J^{[1]})}$ depend on the scheme
and scale $\Lambda$ used to renormalize the color-octet matrix element $\langle \Omega | {\cal O}^{\chi_{cJ}} (^3S_1^{[8]}) | \Omega \rangle$,
which in pNRQCD we identify with the renormalization scheme and scale used for ${\cal E}$.
We renormalize in the $\overline{\rm MS}$ scheme at the scale $\Lambda = m_c$, where, as above, $m_c = 1.5$~GeV is the charm quark mass.\footnote{
  Although from an effective field theory point of view it would make sense to choose for $\Lambda$ a scale close to the soft scale $mv$,
  the specific choice of scale for $\Lambda$ is without consequences here,
  since, according to the renormalization group equation \eqref{eq:corrE_RG}, the $\Lambda$ dependence of $\cal E$ is encoded into an additive constant that may be freely reshuffled
  between $\cal E$ and the short distance coefficients.}

We estimate the uncertainties in the short-distance coefficients to be 30\% of their central values, which account for corrections of relative order $\alpha_{\text{s}}$ or $v^2$ that we neglect.
The variation of the scale $\mu_F$ for the parton distribution functions and of the renormalization scale $\mu_R$ for $\alpha_{\text{s}}$
affects the short-distance coefficients by less than 25\% of the central values.  
We neglect the uncertainty of order $1/N_c^2$ in the wavefunctions (see eq.~\eqref{eq:poperator_result}) compared to other uncertainties.
We use the pNRQCD expressions for the matrix elements in eqs.~\eqref{eq:ldmes_chi}. 
The quarkonium wavefunctions at the origin cancel in the ratio $r_{21}$, which makes $r_{21}$ an attractive observable to extract ${\cal E}$.

Because the transverse momentum $p_T$ of the quarkonium can be much larger than $m_c$ in the kinematical range that we consider, 
the logarithms of $p_T/m_c$ can have a large impact on the perturbative corrections to the short-distance coefficients. 
The resummation of the logarithms of $p_T/m_c$ have been computed in refs.~\cite{Bodwin:2014gia, Bodwin:2015iua} 
at leading logarithmic accuracy at leading power (LP) in the expansion in powers of $m_c/p_T$. 
Following refs.~\cite{Bodwin:2014gia, Bodwin:2015iua}, we use the label  LP+NLO for the short-distance coefficients 
including the additional LP contributions that augment the fixed-order NLO calculations.  
The results in ref.~\cite{Bodwin:2014gia, Bodwin:2015iua} show that the additional LP contributions are numerically significant. 
However, we note also that the large additional LP corrections come mainly from partial contributions  of order $\alpha_\text{s}^2$, 
and may overestimate the size of the corrections to the short-distance coefficients at next-to-next-to-leading order in $\alpha_\text{s}$. 
Therefore, in computing $P$-wave charmonium production cross sections, 
we consider both the fixed-order NLO and LP+NLO calculations of the short-distance coefficients, 
which can provide an estimate of the uncertainty coming from uncalculated corrections  of higher orders in $\alpha_\text{s}$.

\begin{figure}[ht] 
\begin{center}
\includegraphics[width=0.75\textwidth]{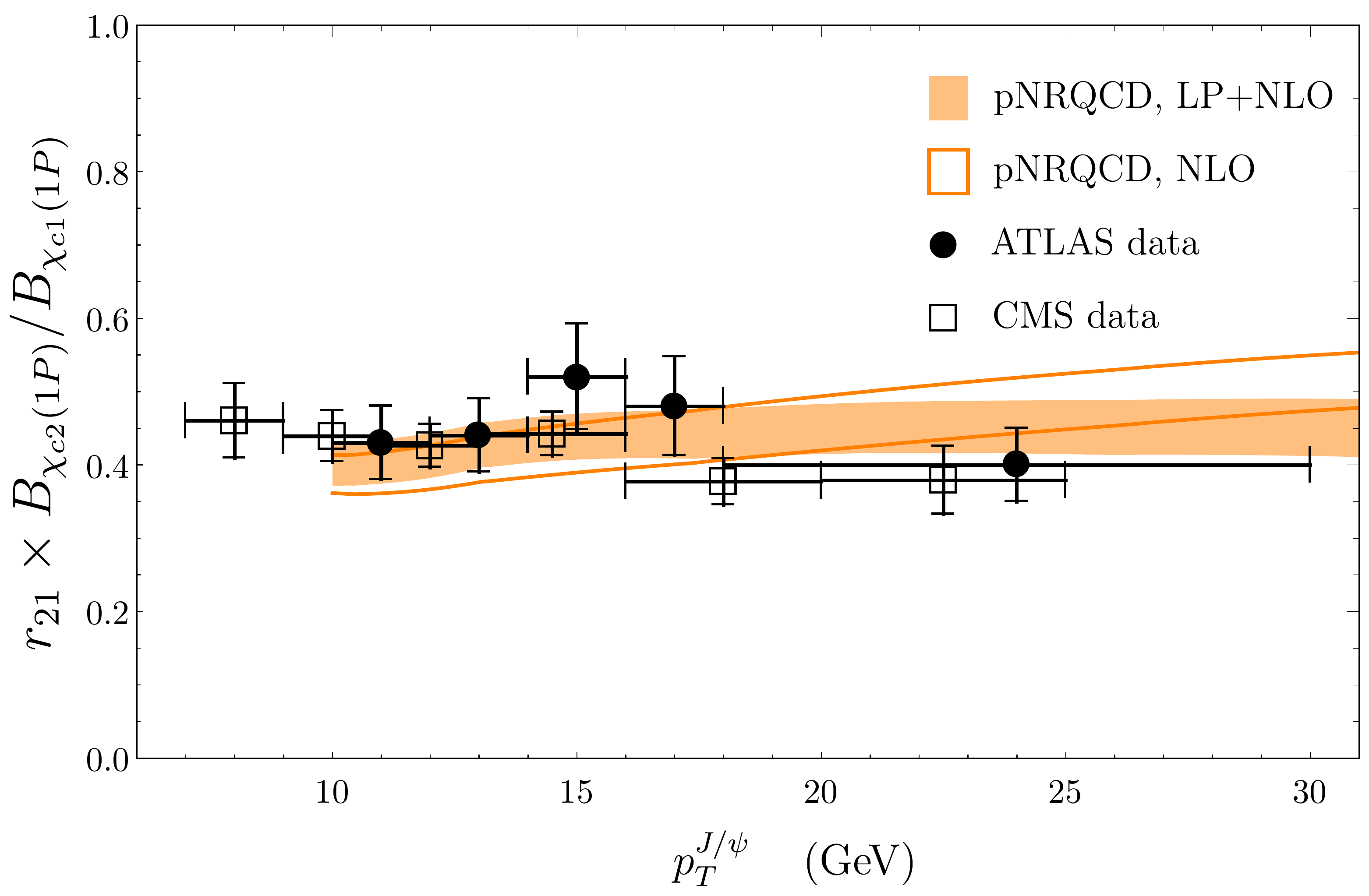}
\caption{\label{fig:chicratio} 
Ratio $r_{21} \times B_{\chi_{c2}(1P)}/B_{\chi_{c1}(1P)}$ of the $\chi_{c2}(1P)$ and $\chi_{c1}(1P)$ differential cross sections at the LHC center of mass energy $\sqrt{s}=7$~TeV and in the rapidity range $|y|<0.75$,
  with fitted ${\cal E}$, compared to CMS~\cite{Chatrchyan:2012ub} and ATLAS~\cite{ATLAS:2014ala} measurements.} 
\end{center}
\end{figure} 

In order to compare to measurements, we compute the values of $r_{21}$ multiplied with $B_{\chi_{c2}(1P)}/ B_{\chi_{c1}(1P)}$,
where $B_{\chi_{cJ}(1P)} = {\rm Br} (\chi_{cJ(1P)} \to J/\psi \gamma) \times {\rm Br} (J/\psi \to \mu^+ \mu^-)$, and Br stands for the branching ratio.  
We compute $B_{\chi_{cJ}(1P)}$ from the PDG values~\cite{Zyla:2020zbs}. 
Since the measurements of $r_{21}$ are given as functions of the transverse momentum $p_T^{J/\psi}$ of the $J/\psi$,
we compute $p_T^{J/\psi}$ from the transverse momentum $p_T$ of the $\chi_{cJ}(1P)$ using 
\begin{equation} 
\label{eq:ptrescale}
p_T^{J/\psi} = \frac{m_{J/\psi}}{m_{\chi_{cJ}(1P)}} p_T, 
\end{equation} 
which is a good approximation since $m_{J/\psi} \approx m_{\chi_{cJ}(1P)}$.  
Corrections to eq.~\eqref{eq:ptrescale} affect the cross section by less than 1\% in the kinematical range that we consider~\cite{Bodwin:2015iua}.
By performing a least-squares fit to the measured values of $r_{21} \times B_{\chi_{c2}(1P)}/ B_{\chi_{c1}(1P)}$ by CMS~\cite{Chatrchyan:2012ub} and ATLAS~\cite{ATLAS:2014ala}, 
we obtain, from fixed-order NLO calculations of the short-distance coefficients,
\begin{equation} 
\label{eq:corrE_value} 
{\cal E}|_{\rm NLO}(\Lambda=1.5\textrm{~GeV}) = 1.17 \pm 0.05,
\end{equation} 
with $\chi^2_{\rm min}/{\rm d.o.f.} = 2.1/10$.  
If we use the LP+NLO expressions  of the short-distance coefficients, we obtain
\begin{equation}
\label{eq:corrE_value2}
{\cal E}|_{\rm LP+NLO}(\Lambda=1.5\textrm{~GeV}) = 4.48 \pm 0.14,
\end{equation}
with $\chi^2_{\rm min}/{\rm d.o.f.} =  1.6/10$.
The difference between the values of ${\cal E}$ in eqs.~\eqref{eq:corrE_value} and~\eqref{eq:corrE_value2} reflects 
the difference between the fixed-order NLO and LP+NLO calculations of the short-distance coefficients.
We show our result for $r_{21}$ compared to ATLAS and CMS data in figure~\ref{fig:chicratio}.  
In the following sections, we will use   both values of ${\cal E}$ 
in eqs.~\eqref{eq:corrE_value} and~\eqref{eq:corrE_value2} to compute cross sections of $\chi_{cJ}$ and $\chi_{bJ}$ states at the LHC. 
More precisely, when computing cross sections of $\chi_{cJ}$, we will use the result in eq.~\eqref{eq:corrE_value} with the fixed-order NLO 
 expressions of the short-distance coefficients, 
 and we will use the result in eq.~\eqref{eq:corrE_value2} with the LP+NLO expressions of the short-distance coefficients.
 When computing cross sections of $\chi_{bJ}$, we will combine the two determinations of ${\cal E}$   into
\begin{equation}
\label{eq:corrE_value3}
{\cal E}(\Lambda=1.5\textrm{~GeV}) = 2.8 \pm 1.7, 
\end{equation}
where the central value is the average of the central values of the determinations in eqs.~\eqref{eq:corrE_value} and~\eqref{eq:corrE_value2}, and the error is such to encompass both determinations.\footnote{
The analysis done in ref.~\cite{Brambilla:2020ojz} contains a normalization error.
After correcting for it, we get a value of $\cal E$ that is compatible with the one in eq.~\eqref{eq:corrE_value2}. 
The value ${\cal E} = 1.94 \pm 0.04$ reported in ref.~\cite{Brambilla:2020ojz} falls nevertheless inside the range given in eq.~\eqref{eq:corrE_value3}.}.

\subsection[Production of $\chi_{cJ}(1P)$]{\boldmath Production of $\chi_{cJ}(1P)$} 
We compute the inclusive production cross sections of $\chi_{cJ}(1P)$ from proton-proton collisions at the LHC
based on the expression of the matrix elements given in eqs.~\eqref{eq:ldmes_chi}. 
We use the same fixed-order NLO and LP+NLO short-distance coefficients from ref.~\cite{Bodwin:2015iua} 
that we used in section~\ref{sec:corrE_determination}, and use the corresponding determinations of $\cal E$ at the scale $\Lambda = 1.5$~GeV given in
eqs.~(\ref{eq:corrE_value}) and (\ref{eq:corrE_value2}) for fixed-order NLO and LP+NLO, respectively.
We determine the value of the charmonium $1P$-state wavefunction at the origin from the two-photon decay rates of the $\chi_{c0}(1P)$ and $\chi_{c2}(1P)$.  
For consistency with our calculation of the cross sections, we use the NRQCD factorization formulas for the decay rates at leading order in $v$, 
while we include order $\alpha_{\text{s}}$ corrections to the short-distance coefficients at the amplitude level.
The pNRQCD expressions for the two-photon widths at leading order in $v$ read~\cite{Bodwin:1994jh, Brambilla:2002nu, Brambilla:2020xod} 
\begin{align} 
\Gamma(\chi_{c0}(1P) \to \gamma \gamma) &= 
\frac{6 \pi e_c^4 \alpha^2}{m_c^4} \left[ 1+ \frac{3 \pi^2-28}{24} C_F \frac{ \alpha_{\text{s}}}{\pi} \right]^2 \frac{3 N_c}{2 \pi} | R^{(0)}{}'(0)|^2, 
\\ 
  \Gamma(\chi_{c2}(1P) \to \gamma \gamma) &=
\frac{8 \pi e_c^4 \alpha^2}{5 m_c^4} \left[ 1- 2 C_F \frac{ \alpha_{\text{s}}}{ \pi} \right]^2 \frac{3 N_c}{2 \pi} | R^{(0)}{}'(0)|^2, 
\end{align} 
where $e_c = 2/3$ is the fractional electric charge of the charm quark, and $\alpha$ is the QED coupling constant.
In the decay rates, we use $\alpha = 1/137$ reflecting the fact that the photons in the final states are on shell,
and use $\alpha_{\text{s}} = 0.282$, which is evaluated at the scale $m_{\chi_{cJ}}/2$.  
We use $m_c = 1.5$~GeV for consistency with the calculation of the cross sections.  
By comparing these formulas with the BESIII measurements of the decay rates~\cite{Ablikim:2012xi},
we obtain $|R^{(0)}{}'(0)|^2 = 0.041$~GeV$^5$ for the $\chi_{c0}(1P)$, and $|R^{(0)}{}'(0)|^2 = 0.073$~GeV$^5$ for the $\chi_{c2}(1P)$.  
For the calculation of the NRQCD matrix elements, we take the average $|R^{(0)}{}'(0)|^2 = 0.057$~GeV$^5$ as the central value.
We attribute to the central value a 30\% uncertainty, which accounts for the uncalculated corrections of relative order $v^2$.  

\begin{figure}[h] 
\begin{center}
\includegraphics[width=0.75\textwidth]{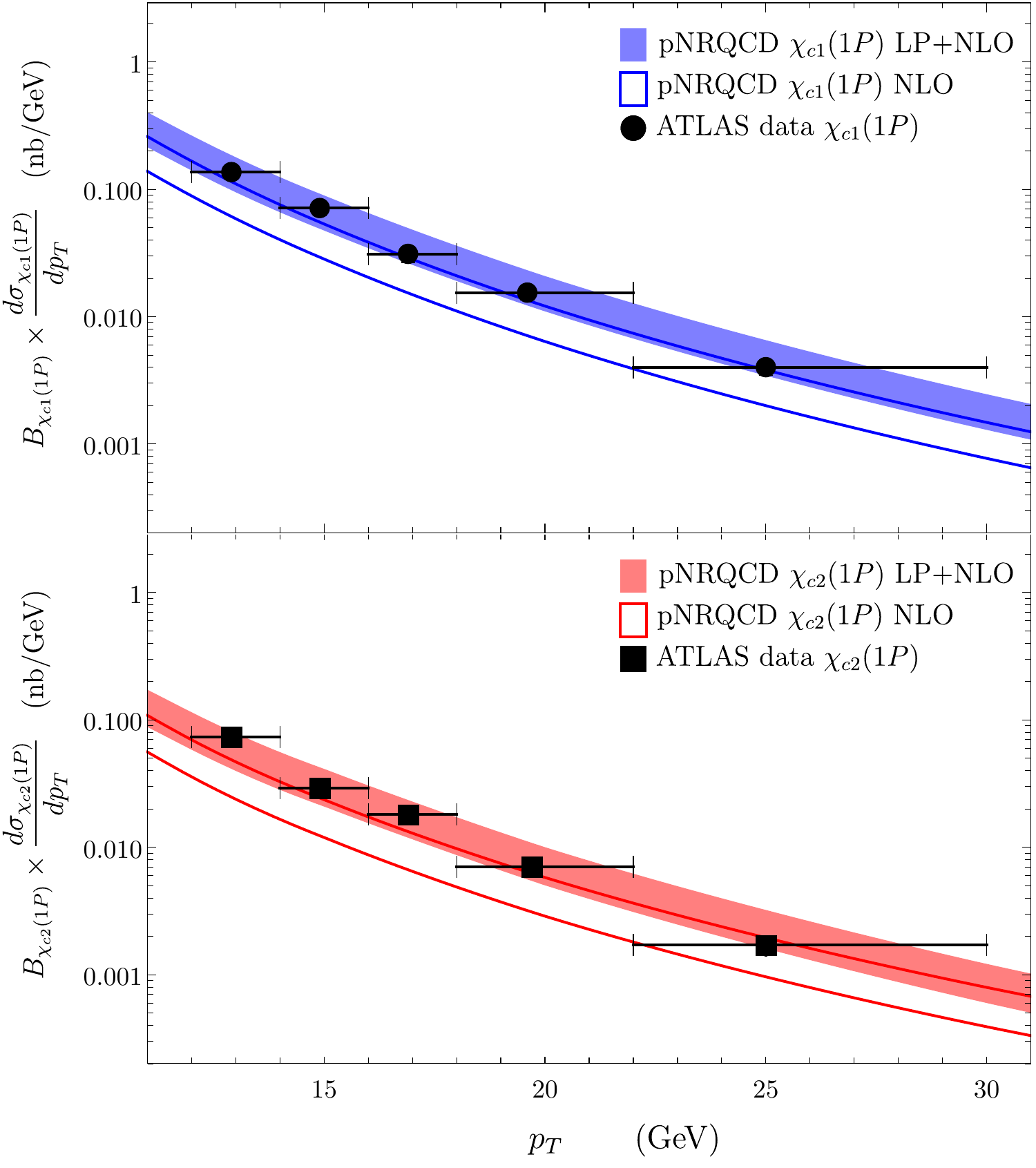}
\caption{\label{fig:chicrate} 
  Production cross sections of the $\chi_{c1}(1P)$ and $\chi_{c2}(1P)$ at the LHC center of mass energy $\sqrt{s}=7$~TeV and in the rapidity range $|y|<0.75$ 
  compared with ATLAS measurements~\cite{ATLAS:2014ala}. } 
\end{center}
\end{figure} 

We note that, while the order-$\alpha_\text{s}^2$ corrections to the two-photon widths have been computed in ref.~\cite{Sang:2015uxg}, 
these corrections depend strongly on the scheme and the scale associated with the renormalization of the wavefunctions at the origin  
(see, for example, refs.~\cite{Chung:2020zqc, Chung:2021efj}). 
The strong dependence on the scheme and scale will not cancel in production rates 
unless the short-distance coefficients for cross sections are also computed to the same accuracy in $\alpha_\text{s}$. 
Since the short-distance coefficients for heavy quarkonium production are currently known at NLO accuracy in $\alpha_\text{s}$, 
we neglect the order-$\alpha_\text{s}^2$ corrections to the two-photon widths. 

The results for the  $\chi_{c1}(1P)$ and $\chi_{c2}(1P)$ cross sections are shown in figure~\ref{fig:chicrate} against ATLAS data.  
The fixed-order NLO and LP+NLO calculations of the short-distance coefficients give results for the cross sections that are compatible within uncertainties. 
We note that the NLO results for the cross sections are systematically lower than the ATLAS data in particular at low $p_T$, while the LP+NLO results agree well with the measurements.

\subsection[Polarization of $\chi_{cJ}(1P)$]{\boldmath Polarization of $\chi_{cJ}(1P)$} 
In the case of polarized cross sections, the non-isotropic part of ${\cal E}^{ij}$ can in principle contribute to the color-octet matrix elements,
and, if such contribution is nonvanishing, the color-octet matrix elements can acquire a dependence on the arbitrary direction of the gauge-completion Wilson lines.  
For the universality of the NRQCD matrix elements to be valid also for the case of polarized cross sections, such non-isotropic contributions should vanish in the NRQCD matrix elements.
This has not been proved.
However, it is common practice in the phenomenological literature to assume that the polarized color-octet LDMEs  
do not depend on the direction of the gauge-completion Wilson lines, and, therefore, do not depend on the polarization $\lambda$
(see for instance refs.~\cite{Butenschoen:2012px, Butenschoen:2012qr,Gong:2012ug, Shao:2014fca, Bodwin:2015iua, Faccioli:2020rue}).
This amounts at assuming that the polarized color-octet LDMEs
$\langle \Omega | \chi^\dag \sigma^i T^a \psi \Phi_\ell^{\dag ab} (0) {\cal P}_{{\chi_{QJ}}(\lambda, \bm{P}=\bm{0})} \Phi_\ell^{bc} (0)$ $ \psi^\dag \sigma^i T^c \chi| \Omega \rangle$
can be related for any polarization $\lambda$ of the $\chi_{QJ}$ state to the polarization-summed color-octet LDME $\langle \Omega | {\cal O}^{\chi_{QJ}}({}^3S_1^{[8]}) | \Omega \rangle$,
which is independent of the direction of the gauge-completion Wilson lines, through the relation
\begin{equation} 
\label{eq:co_ldme_chipol} 
\langle \Omega | {\cal O}^{\chi_{QJ}}({}^3S_1^{[8]}) | \Omega \rangle 
= (2 J+1) \times \langle \Omega | \chi^\dag \sigma^i T^a \psi \Phi_\ell^{\dag ab} (0) {\cal P}_{{\chi_{QJ}}(\lambda, \bm{P}=\bm{0})} \Phi_\ell^{bc} (0) \psi^\dag \sigma^i T^c \chi | \Omega \rangle,
\end{equation} 
and similarly for $h_Q$:
\begin{equation} 
\label{eq:co_ldme_hqpol}
\langle \Omega | {\cal O}^{h_{Q}}({}^1S_0^{[8]}) | \Omega \rangle
= 3 \times \langle \Omega | \chi^\dag T^a \psi \Phi_\ell^{\dag ab} (0) {\cal P}_{{h_{Q}}(\lambda, \bm{P}=\bm{0})} \Phi_\ell^{bc} (0) \psi^\dag T^c \chi | \Omega \rangle.  
\end{equation} 
These relations are  consequences of  the heavy quark spin symmetry and the assumption of the universality of the NRQCD matrix elements for polarized quarkonia
in the right-hand sides of eqs.~\eqref{eq:co_ldme_chipol} and \eqref{eq:co_ldme_hqpol}.  
Similar relations hold for the color-singlet LDMEs:
\begin{subequations} 
\label{eq:cs_ldme_chipol} 
\begin{align} 
  \langle \Omega | {\cal O}^{\chi_{Q1}}({}^3P_1^{[1]}) | \Omega \rangle
&= 3 \times \langle \Omega | \frac{1}{2} \chi^\dag \left(-\frac{i}{2} \overleftrightarrow{\bm{D}} \times \bm{\sigma} \right)^i \psi {\cal P}_{\chi_{Q1}(\lambda, \bm{P}=\bm{0})} \psi^\dag 
\left(-\frac{i}{2} \overleftrightarrow{\bm{D}} \times \bm{\sigma} \right)^i \chi | \Omega \rangle, 
\\ 
  \langle \Omega | {\cal O}^{\chi_{Q2}}({}^3P_2^{[1]}) | \Omega \rangle
&= 5 \times \langle \Omega | \chi^\dag \left(-\frac{i}{2} \overleftrightarrow{{D}}^{(i} {\sigma}^{j)} \right) \psi {\cal P}_{\chi_{Q2}(\lambda, \bm{P}=\bm{0})} \psi^\dag
\left(-\frac{i}{2} \overleftrightarrow{{D}}^{(i} {\sigma}^{j)} \right) \chi  | \Omega \rangle, 
\\ 
\langle \Omega | {\cal O}^{h_Q}({}^1P_1^{[1]}) | \Omega \rangle
&= 3 \times \langle \Omega | \chi^\dag \left(-\frac{i}{2} \overleftrightarrow{D}^i \right) \psi {\cal P}_{{h_Q}(\lambda,\bm{P}=\bm{0})} \psi^\dag
\left(-\frac{i}{2} \overleftrightarrow{D}^i \right) \chi | \Omega \rangle.
\end{align} 
\end{subequations} 
In this case, however, they are not assumptions, but follow from the vacuum-saturation approximation and rotational symmetry.  

\begin{figure}[ht] 
\begin{center}
\includegraphics[width=0.75\textwidth]{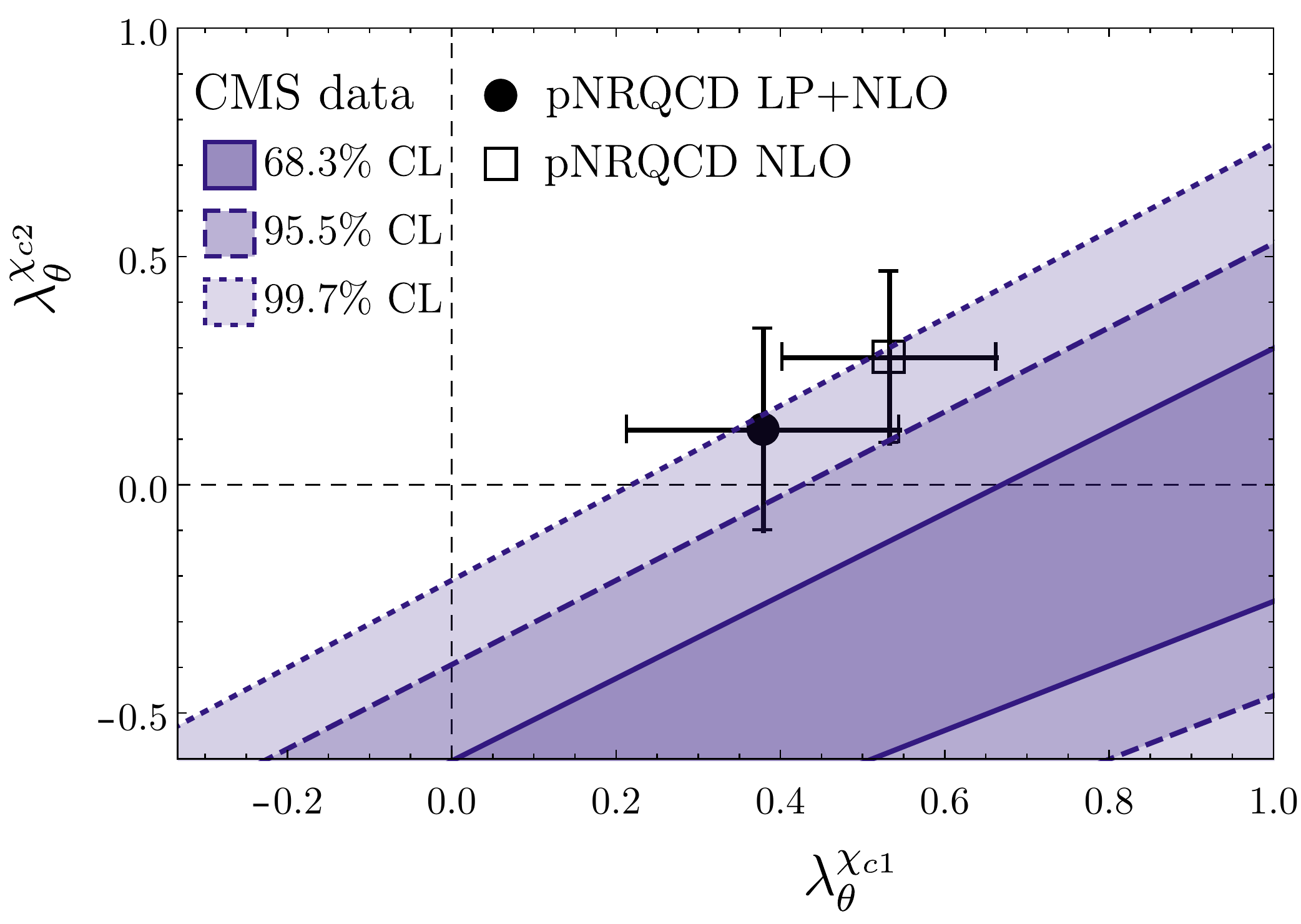} 
\caption{\label{fig:chicpol} 
  Polarization of $\chi_{c2}(1P)$ and $\chi_{c1}(1P)$ at the LHC center of mass energy $\sqrt{s}=7$~TeV and in the rapidity range $|y|<0.75$,
  averaged over the range 8~GeV$<p_T^{J/\psi}<$30~GeV, compared with experimental constraints from CMS~\cite{Sirunyan:2019apc}.} 
\end{center}
\end{figure} 

Under the universality assumption of the polarized color-octet LDMEs, we can compute the polarization of $\chi_{c1}(1P)$ and $\chi_{c2}(1P)$ produced at the LHC.
The polarization parameters $\lambda_\theta^{\chi_{c1}}$ and $\lambda_\theta^{\chi_{c2}}$ are defined by 
\begin{equation} 
\lambda_\theta^{\chi_{cJ}} = \frac{1 - 3 \xi_{\chi_{cJ}}}{1 + \xi_{\chi_{cJ}}}, 
\end{equation} 
where $\xi_{\chi_{cJ}}$ is the fraction of $J/\psi$ produced with longitudinal polarization from decays of $\chi_{cJ}(1P)$.  
We use the hadron helicity frame to define the spin quantization axis of the $J/\psi$.
The polarized cross sections can be computed by using the short-distance coefficients from ref.~\cite{Bodwin:2015iua} for the polarized production of $\chi_{cJ}(1P)$, 
and eq.~\eqref{eq:co_ldme_chipol} to relate the color-octet LDMEs for polarized and polarization summed $\chi_{cJ}(1P)$ states.
Equation~\eqref{eq:co_ldme_chipol} relies on the assumption that the NRQCD matrix elements are universal for polarized cross sections.  
Our results for $\lambda_\theta^{\chi_{c1}}$ and $\lambda_\theta^{\chi_{c2}}$ are shown in figure~\ref{fig:chicpol} against experimental constraints coming from the CMS data~\cite{Sirunyan:2019apc}.  

It has been pointed out in refs.~\cite{Faccioli:2010kd, Faccioli:2014cqa,Faccioli:2017hym, Faccioli:2020rue} that modifications to the central values of $\chi_{cJ}$ cross section measurements
in refs.~\cite{Chatrchyan:2012ub, ATLAS:2014ala} due to polarizations of $\chi_{c1}$ and $\chi_{c2}$ may be important.
The central values in refs.~\cite{Chatrchyan:2012ub, ATLAS:2014ala} have been obtained by assuming isotropic decay angular distributions,
while the scale factors that modify the central cross-section values are listed for extreme polarization scenarios.
We compute the scale factors by using our results for $\lambda_\theta^{\chi_{c1}}$ and $\lambda_\theta^{\chi_{c2}}$ in figure~\ref{fig:chicpol}, 
and by assuming that the scale factor depends smoothly on polarization. 
The scale factors for $r_{21}$ that we obtain are about 0.95 for CMS data~\cite{Chatrchyan:2012ub}, and about 0.89 for ATLAS data~\cite{ATLAS:2014ala}.
These scale factors have mild effects  on our phenomenological determinations of $\cal E$, 
and slightly reduce the central values by less than the uncertainties in $\cal E$. 
In ref.~\cite{Faccioli:2020rue}, a smaller scale factor of 0.85 has been obtained by using a data-driven analysis.
In this case, the values of ${\cal E}$ determined from $r_{21}$ reduce by slightly more than the estimated uncertainties. 
Nonetheless, the changes in the central values of ${\cal E}$ have negligible effect on our phenomenological results for $P$-wave charmonia and bottomonia.

\subsection[Production of $\chi_{bJ}(nP)$]{\boldmath Production of $\chi_{bJ}(nP)$} 
\label{sec:chibprod}
Owing to the universality of the correlator $\cal E$, 
we can compute in pNRQCD, at leading order in $v$, inclusive production cross sections of $\chi_{bJ}(nP)$ from proton-proton collisions at the LHC without having to fit new octet LDMEs.
We first consider the ratio $r_{21}$ for the differential cross sections of $\chi_{b2}(nP)$ and $\chi_{b1}(nP)$ states.
We compute the short-distance coefficients at next-to-leading order accuracy using the FDCHQHP package~\cite{Wan:2014vka}.
We use the bottom quark mass $m_b= 4.75$~GeV, CTEQ6M parton distribution functions at the scale $\mu_F = \sqrt{p_T^2+4 m_b^2}$,
and compute $\alpha_{\text{s}}$ at the same scale running at two loops with $n_f = 5$ light quark flavors and $\Lambda_{\rm QCD}^{(5)} = 226$~MeV.  
Since the range of $p_T$ for the $\chi_{bJ}(nP)$ that we consider is not too large compared to the mass of the $\chi_{bJ}(nP)$ states, we do not resum logarithms of $p_T/m_b$.  
We take the renormalization scale $\Lambda$ of the color-octet matrix element $\langle \Omega | {\cal O}^{\chi_{bJ}(nP)} (^3S_1^{[8]}) | \Omega \rangle$,
which in pNRQCD is the renormalization scale of $\cal E$, to be $m_b$.  
We compute ${\cal E}$ at the scale $4.75$~GeV by using the one-loop renormalization-group-improved  formula 
\begin{equation} 
{\cal E} (m_b) = {\cal E} (m_c) + \frac{24 C_F}{\beta_0} \log \frac{\alpha_{\text{s}}(m_c)}{\alpha_{\text{s}}(m_b)}, 
\end{equation} 
where $\beta_0 = 11 N_c/3-2 n_f/3$ and 
${\cal E} (m_c)$ is given in
eq.~\eqref{eq:corrE_value3}.
We take into account the uncertainty in ${\cal E}$, and estimate the uncertainties from uncalculated corrections of order $v^2$ to be 10\% of the central values.  

\begin{figure}[ht] 
\begin{center}
\includegraphics[width=0.75\textwidth]{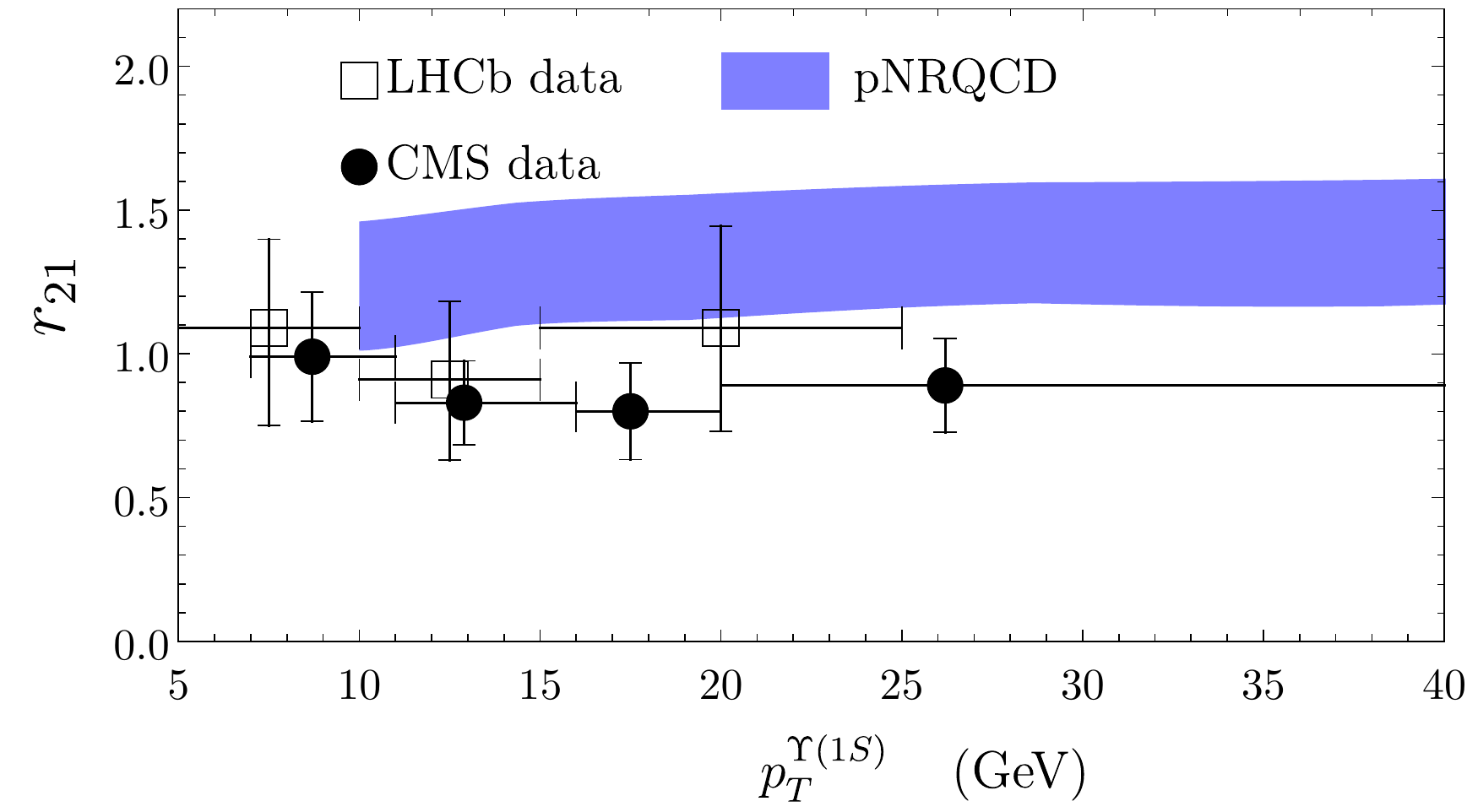}
\caption{\label{fig:chibratio} 
  Ratio of $\chi_{b2}(1P)$ and $\chi_{b1}(1P)$ differential cross sections at the LHC center of mass energy $\sqrt{s}=7$~TeV and in the rapidity range $2 < y < 4.5$ 
  compared with LHCb~\cite{Aaij:2014hla} and CMS~\cite{Khachatryan:2014ofa} measurements.} 
\end{center}
\end{figure} 

Our result for $r_{21}$ in the case of $\chi_{bJ}(1P)$ states is shown in figure~\ref{fig:chibratio} against LHCb~\cite{Aaij:2014hla} and CMS~\cite{Khachatryan:2014ofa} data.  
The result is given as a function of the transverse momentum $p_T^{\Upsilon(1S)}$ of the $\Upsilon(1S)$ coming from the decay of the $\chi_{bJ}(1P)$,
which we relate to the transverse momentum $p_T$ of the $\chi_{bJ}(1P)$ by 
\begin{equation} 
p_T^{\Upsilon(1S)} = \frac{m_{\Upsilon(1S)}}{m_{\chi_{bJ}(1P)}} p_T.  
\end{equation} 
In figure~\ref{fig:chibratio}, we did not include the feeddown contributions, which come dominantly from decays of $\Upsilon(2S)$ and $\Upsilon(3S)$.
Since the branching ratios into $\chi_{b1}(1P)$ and $\chi_{b2}(1P)$ are almost the same, and our results for $r_{21}$ is close to unity,
we expect our results for $r_{21}$ to be almost unchanged by the inclusion of the feeddown contributions.  

In pNRQCD, the wavefunction dependence factorizes in the cross sections, see eqs. \eqref{eq:fac_pwave_pNRQCD},
and cancels at leading order in $v$ when considering the ratio of states with the same principal quantum number.
Moreover, the ratio $r_{21}$ for the $\chi_{bJ}(1P)$ states in figure~\ref{fig:chibratio} is almost independent of $p_T$.
It is, therefore, a specific prediction of (leading order) pNRQCD that the ratio $r_{21}$ is independent of the principal quantum number,
i.e. that the ratios for the $\chi_{bJ}(2P)$ and $\chi_{bJ}(3P)$ states is expected to be the same as the one shown in figure~\ref{fig:chibratio} for the $\chi_{bJ}(1P)$ states.

For the computation of the production rates of $\chi_{bJ}(nP)$ we need the values of the $P$-wave bottomonium wavefunctions at the origin.
Since the decay widths of the $\chi_{bJ}(nP)$ states are in general not well known, differently from the $\chi_{cJ}(1P)$ states,
we take the central values of the wavefunctions at the origin to be the average of the potential-model calculations considered in ref.~\cite{Brambilla:2020xod}.
They are
\begin{subequations} 
\begin{align} 
|R^{(0)}_{1P}{}' (0)|^2 &= 1.47~{\rm GeV}^5,\\ 
|R^{(0)}_{2P}{}' (0)|^2 &= 1.74~{\rm GeV}^5,\\ 
|R^{(0)}_{3P}{}' (0)|^2 &= 1.92~{\rm GeV}^5.  
\end{align} 
\end{subequations} 
We take the uncertainties in the wavefunctions at the origin to be 10\% of the central values, which account for the uncalculated corrections of order $v^2$.  

\begin{figure}[ht]
\begin{center}
\includegraphics[width=0.75\textwidth]{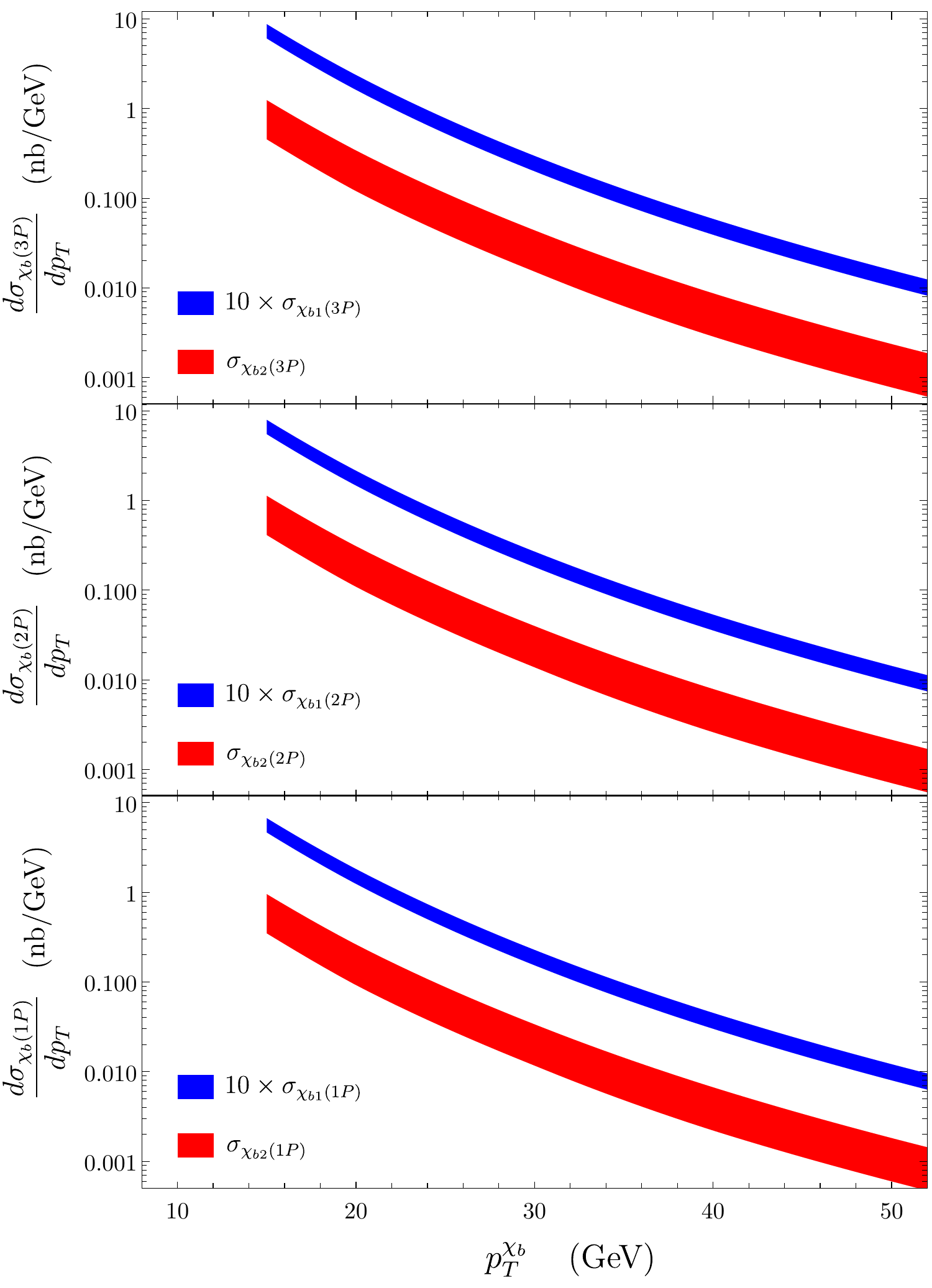}
\caption{\label{fig:chibabsrate}
Differential production cross sections of the $\chi_{b1} (nP)$ and $\chi_{b2} (nP)$ ($n=1$, 2, and 3) at the LHC center of mass energy $\sqrt{s}=7$~TeV and in the rapidity range $2 < y < 4.5$.
The  $\chi_{b1} (nP)$ cross sections have been multiplied by 10 to distinguish them from the $\chi_{b2} (nP)$ cross sections on the logarithmic scale.
}
\end{center}
\end{figure}

While the $p_T$-dependent absolute production rates of $\chi_{bJ}$ states at the LHC have not been reported yet,
we compute cross sections of $\chi_{b1}(nP)$ and $\chi_{b2} (nP)$ ($n=1$, 2, and 3) at the LHC with $\sqrt{s}=7$~TeV in the rapidity range $2 < y < 4.5$.
Our results for the differential cross sections are shown in figure~\ref{fig:chibabsrate}.
Once measurements become available, our predictions could be compared with data.

To compare with data we need to compute the feeddown fraction $R_{\Upsilon(n'S)}^{\chi_{b}(nP)}$,
which is defined as the fraction of $\Upsilon(n'S)$ produced from decays of $\chi_{b1}(nP)$ and $\chi_{b2}(nP)$.  
We compute $R_{\Upsilon(n'S)}^{\chi_{b}(nP)}$ by using the formula 
\begin{equation} 
R_{\Upsilon(n'S)}^{\chi_{b}(nP)} = \frac{\sum_{J=1,2} {\rm Br}({\chi_{bJ} (nP) \to \Upsilon(n'S)+\gamma})  \times \sigma_{\chi_{bJ}(nP)} } { \sigma_{\Upsilon(n'S)}}, 
\end{equation} 
where the inclusive cross sections $\sigma_{\chi_{bJ}(nP)}$ and $\sigma_{\Upsilon(n'S)}$ are the sum of the direct production rate and the feeddown contribution.  
We neglect the contributions from $\chi_{b0}(nP)$, because the branching ratios ${\rm Br}({\chi_{b0} (nP) \to \Upsilon(n'S)+\gamma})$ are small,
while the cross sections $\sigma_{\chi_{b0} (nP)}$ are similar in size compared to $\sigma_{\chi_{b1}(nP)}$ or $\sigma_{\chi_{b2} (nP)}$.  
The feeddown contribution to the $\Upsilon(n'S)$ production rate is given by 
\begin{align} 
\sigma_{\Upsilon(n'S)}^{\rm feeddown} 
&= \sum_{n \geq n'} \sum_{J=1,2} {\rm Br}({\chi_{bJ} (nP) \to \Upsilon(n'S)+\gamma}) \times \sigma_{\chi_{bJ}(nP)}
\nonumber \\ 
& \hspace{5ex} + 
\sum_{n>n'} {\rm Br} (\Upsilon(nS) \to \Upsilon(n'S) + X) \times \sigma_{\Upsilon(nS)}, 
\end{align} 
where we truncate the sum at $n=3$.
In the production rate of $\Upsilon(3S)$, we only consider the feeddown contribution from decays of $\chi_b(3P)$.  
In the production rate of $\Upsilon(2S)$, we consider the feeddown contributions from decays of $\chi_b(3P)$ and $\chi_b(2P)$, and also from decays of $\Upsilon(3S)$.  
For the $\Upsilon(1S)$ production, we consider the feeddowns from decays of $\chi_b(3P)$, $\chi_b(2P)$, $\chi_b(1P)$, $\Upsilon(3S)$ and $\Upsilon(2S)$ states.  
For the production of the $\chi_{bJ}(3P)$ state we neglect the contribution from feeddowns,
while the feeddown contribution for the $\chi_{bJ}(nP)$ production rate for $n=1$, $2$ is given by 
\begin{equation} 
\sigma_{\chi_{bJ}(nP)}^{\rm feeddown} = \sum_{n'>n} {\rm Br}(\Upsilon(n'S) \to \chi_{bJ}(nP) + \gamma) \times \sigma_{\Upsilon(n'S)}, 
\end{equation} 
where we truncate the sum again at $n'=3$.  
While we compute the direct production cross sections of $\chi_{bJ} (nP)$ in pNRQCD from eq.~\eqref{eq:fac_pwave_pNRQCDb} (the result has been shown in figure~\ref{fig:chibabsrate}),
we compute the direct production rates of $\Upsilon(n'S)$ from the NRQCD factorization formula:  
\begin{align} 
\sigma_{\Upsilon(n'S)}^{\rm direct} 
&= 
\sigma_{Q \bar Q (^3S_1^{[1]})} \langle \Omega | {\cal O}^{\Upsilon(n'S)}(^3S_1^{[1]}) | \Omega \rangle + \sigma_{Q \bar Q (^3S_1^{[8]})} \langle \Omega | {\cal O}^{\Upsilon(n'S)}(^3S_1^{[8]}) | \Omega \rangle 
\nonumber \\ 
& \hspace{5ex} 
+ \sigma_{Q \bar Q (^1S_0^{[8]})} \langle \Omega | {\cal O}^{\Upsilon(n'S)}(^1S_0^{[8]}) | \Omega \rangle + \sigma_{Q \bar Q (^3P_J^{[8]})} \langle \Omega | {\cal O}^{\Upsilon(n'S)}(^3P_0^{[8]}) | \Omega \rangle, 
\end{align} 
which is truncated at relative order $v^4$.
The NRQCD operators for production of $\Upsilon(nS)$ are defined by 
\begin{subequations} 
\begin{align} 
{\cal O}^{\Upsilon(n'S)}({}^3S_1^{[1]}) &= \sum_{\lambda} \chi^\dag \sigma^i \psi {\cal P}_{{\Upsilon(n'S)}(\lambda,\bm{P}=\bm{0})} \psi^\dag \sigma^i \chi, 
\\ 
{\cal O}^{\Upsilon(n'S)}({}^3S_1^{[8]}) &= \sum_{\lambda} \chi^\dag \sigma^i T^a \psi \Phi_\ell^{\dag ab} (0) {\cal P}_{{\Upsilon(n'S)}(\lambda, \bm{P}=\bm{0})} \Phi_\ell^{bc} (0) \psi^\dag \sigma^i T^c \chi, 
\\ 
{\cal O}^{\Upsilon(n'S)}({}^1S_0^{[8]}) &= \sum_{\lambda} \chi^\dag T^a \psi \Phi_\ell^{\dag ab} (0) {\cal P}_{{\Upsilon(n'S)}(\lambda, \bm{P}=\bm{0})} \Phi_\ell^{bc} (0) \psi^\dag T^c \chi, 
\\ 
{\cal O}^{\Upsilon(n'S)}({}^3P_0^{[8]}) &= \sum_\lambda \frac{1}{3} \chi^\dag \left(\!-\frac{i}{2} \overleftrightarrow{\bm{D}} \cdot \bm{\sigma} \!\right) T^a \Phi_\ell^{\dag ab} (0) 
\psi {\cal P}_{\Upsilon(n'S)(\lambda,\bm{P}=\bm{0})} \Phi_\ell^{bc} (0) \psi^\dag \left(\!-\frac{i}{2} \overleftrightarrow{\bm{D}} \cdot \bm{\sigma} \!\right) T^c \chi.  
\end{align} 
\end{subequations} 
We have used the heavy quark spin symmetry to relate the LDMEs $\langle \Omega | {\cal O}^{\Upsilon(n'S)}({}^3P_J^{[8]}) | \Omega \rangle$ to the $J=0$ one,
where the operators ${\cal O}^{\cal Q}({}^3P_J^{[8]})$ are obtained from ${\cal O}^{\cal Q}({}^3P_J^{[1]})$ by inserting color matrices
and gauge-completion Wilson lines between the quark and antiquark fields.
As is commonly done in phenomenological studies of $S$-wave heavy quarkonium production,
we neglect contributions from color-singlet matrix elements of relative orders $v^2$ and higher that are obtained by inserting powers of $\overleftrightarrow{\bm{D}}^2$
between the quark and antiquark fields, because their contributions to the cross section turn out to be numerically small. 
Since the color-octet matrix elements for the $\Upsilon(n'S)$ states have not been computed from first principles,
we take the values that were determined in ref.~\cite{Han:2014kxa} from fits to measured cross sections at the LHC.  
We take the measured values of the branching ratios from ref.~\cite{Zyla:2020zbs}. 
Since the branching ratios of the $\chi_{bJ}(3P)$ states have not been measured yet, we use the theoretical prediction of ref.~\cite{Han:2014kxa}.  
For the branching ratios ${\rm Br} (\Upsilon(3S) \to \Upsilon(1S) + X)$ and ${\rm Br} (\Upsilon(2S) \to \Upsilon(1S) + X)$,
we take the sum of the branching ratios into $\Upsilon(1S) \pi^+ \pi^-$ and $\Upsilon(1S) \pi^0 \pi^0$.  

\begin{figure}[ht] 
\begin{center}
\includegraphics[width=0.75\textwidth]{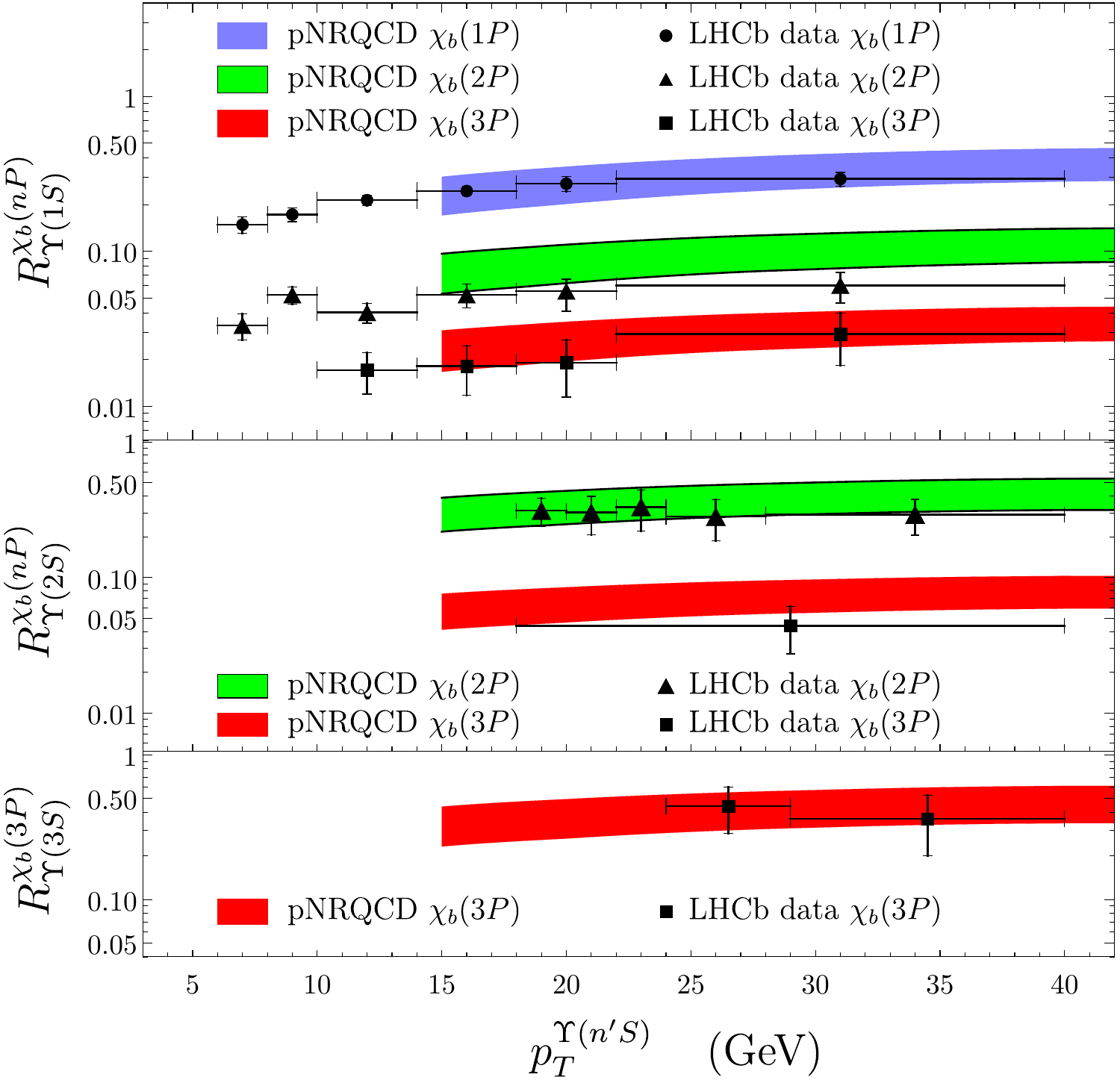} 
\caption{\label{fig:chibrate} 
  Feeddown fractions $R_{\Upsilon(n'S)}^{\chi_b(nP)}$ at the LHC center of mass energy $\sqrt{s}=7$~TeV and in the rapidity range $2 < y < 4.5$ compared with LHCb data~\cite{Aaij:2014caa}.} 
\end{center}
\end{figure} 

Our pNRQCD results for the feeddown fractions $R_{\Upsilon(n'S)}^{\chi_b(nP)}$ are shown in figure~\ref{fig:chibrate} against LHCb data~\cite{Aaij:2014caa}.  
Compared to the previous pNRQCD results in ref.~\cite{Brambilla:2020ojz}, 
the results here have larger uncertainties because we consider a wider range of values  for $\cal E$, which reflect the uncertainty 
 due to  unknown corrections of higher orders in $\alpha_\text{s}$ in the short-distance coefficients.
Also, 
 the results here include feeddown contributions to $\chi_b(2P)$ and $\chi_b(1P)$ cross sections,
which increase slightly the fractions $R_{\Upsilon(n'S)}^{\chi_b(2P)}$ and $R_{\Upsilon(n'S)}^{\chi_b(1P)}$, and improve the agreement with the LHCb data.

\section{Conclusions} 
\label{sec:conclusion} 
In this work, we have developed a formalism for computing the NRQCD long-distance matrix elements for the inclusive production of heavy quarkonia, based on pNRQCD.
Our formalism applies to strongly coupled quarkonia, which include excited charmonium and bottomonium states.
We obtain expressions of the LDMEs that are given by products of quarkonium wavefunctions at the origin and universal gluonic correlators, analogously to the pNRQCD expressions of decay LDMEs.
The computation of the production LDMEs in this paper is valid up to corrections of relative order $1/N_c^2$ and $v^2$.  
Corrections of higher orders in $v$ come from higher order corrections in the pNRQCD expressions of the NRQCD LDMEs, as well as from corrections to the quarkonium wavefunctions.

Based on the general formalism that we have developed in section~\ref{sec:ldmes},
we have computed NRQCD LDMEs for hadroproduction of $P$-wave heavy quarkonia at leading nonvanishing order in $v$ in section~\ref{sec:pwave-th}.
For the case of color-singlet LDMEs, we reproduce the known result from the vacuum-saturation approximation
where the production LDMEs and decay LDMEs are both given by the modulus square of the derivative of the quarkonium wavefunctions at the origin.
For the color-octet LDMEs, we obtain expressions that are similar to the color-octet decay LDMEs,
except that they depend on a gluonic correlator $\cal E$ that is different from the one found for decay LDMEs,
due to a different field arrangement and the gauge-completion Wilson lines that are necessary to ensure the gauge invariance of the color-octet LDMEs.  
The results confirm our previous calculation of the $P$-wave LDMEs in ref.~\cite{Brambilla:2020ojz}, where we assumed heavy-quark spin symmetry to reduce the number of independent LDMEs.
In this work, we explicitly confirm the validity of the heavy-quark spin symmetry for $P$-wave LDMEs at leading nonvanishing order in $v$.  

Our results for the color-octet $P$-wave LDMEs help identify the infrared behavior, which is necessary in testing the validity of the NRQCD factorization in heavy quarkonium production.
Since the color-octet $P$-wave LDMEs are given by products of the modulus square of the derivatives of the wavefunctions at the origin
and the gluonic correlator $\cal E$, defined in eq.~\eqref{eq:ecorrelator}, 
they have additional infrared divergences that come from $\cal E$ compared to the color-singlet $P$-wave LDMEs.
These are consistent with the infrared factor found computing the gluon fragmentation function in ref.~\cite{Nayak:2005rt},
which, in turn, agrees with the explicit calculation of the infrared divergences in the color-octet LDMEs in ref.~\cite{Nayak:2006fm}.
Therefore, our results are consistent with the known infrared behavior of the NRQCD LDMEs.
Furthermore, our expressions for the color-octet $P$-wave LDMEs for polarization-summed cross sections are independent of the direction of the gauge-completion Wilson lines.  
This implies that the color-octet $P$-wave LDMEs are process independent, which is a necessary condition for the validity of the NRQCD factorization.  
Hence, our results for the $P$-wave LDMEs support the validity of the NRQCD factorization in heavy quarkonium production for polarization-summed cross sections.  

We have computed production cross sections of $\chi_{cJ}$ and $\chi_{bJ}$ states at the LHC for $J=1$, $2$
based on our calculations of the $P$-wave LDMEs and the NRQCD short-distance coefficients that are available through next-to-leading order accuracy in $\alpha_{\text{s}}$.  
Since a lattice QCD determination of the correlator $\cal E$ is not available, 
we determine $\cal E$ by comparing the theoretical expression of the differential cross section ratio $(d\sigma_{\chi_{c2}(1P)}/dp_T)/(d\sigma_{\chi_{c1}(1P)}/dp_T)$
with measurements at the LHC by CMS~\cite{Chatrchyan:2012ub} and ATLAS~\cite{ATLAS:2014ala}.  
This allows a phenomenological determination of $\cal E$ that does not depend on the value of the wavefunction at the origin.
This determination improves our previous determination in ref.~\cite{Brambilla:2020ojz} where absolute cross section measurements were used.
After fixing the value of the derivative of the $P$-wave charmonium wavefunction at the origin on the measured two-photon decay rates, 
we have computed the cross sections of $\chi_{c1}(1P)$ and $\chi_{c2}(1P)$ at the LHC, which agree well with ATLAS measurements~\cite{ATLAS:2014ala}.
Our results for the $\chi_{c1}(1P)$ and $\chi_{c2}(1P)$ cross sections are consistent with our previous results in ref.~\cite{Brambilla:2020ojz}.  
We have also computed the polarizations of $\chi_{c1}(1P)$ and $\chi_{c2}(1P)$ at the LHC under the assumption that the NRQCD factorization holds for polarized cross sections.
Our results are consistent with experimental constraints from CMS~\cite{Sirunyan:2019apc}.
The universality of the correlator $\cal E$ allows to compute $P$-wave bottomonium cross sections without having to fit any new octet LDME.
From a phenomenological point of view, this is the most relevant gain in the pNRQCD approach.
In the bottomonium case, we have computed the differential cross section ratio $(d\sigma_{\chi_{b2}(1P)}/dp_T)/(d\sigma_{\chi_{b1}(1P)}/dp_T)$
and the feeddown fractions $R_{\Upsilon(n'S)}^{\chi_b(nP)}$ at the LHC, for which measurements are available.  
We find good agreements with data~\cite{Aaij:2014hla,Khachatryan:2014ofa, Aaij:2014caa}.
We have improved our results for the feeddown fractions $R_{\Upsilon(n'S)}^{\chi_b(nP)}$ compared to our previous results in ref.~\cite{Brambilla:2020ojz}
by considering also the contributions from cascade decays. This results in a better agreement with data.
Finally, we made predictions for absolute production rates of $\chi_{b1}(nP)$ and $\chi_{b2}(nP)$ ($n=1$, 2, and 3) at the LHC, which may be compared with data once measurements become available. 

The phenomenological results in this work are based on the determination of the correlator $\cal E$.
We have obtained it from measured cross sections of charmonia.  
It would be desirable, however, to have a lattice QCD determination of this nonperturbative quantity.
Apart from its phenomenological relevance, a lattice QCD determination of $\cal E$ would also provide a nontrivial test of QCD in the nonperturbative regime.
While in the charmonium sector, the derivative of the wavefunctions at the origin can be extracted from two-photon decay data,
this is not yet possible in the bottomonium sector for lack of data.
In the meantime, relying on models is a major source of uncertainty, and reducing this uncertainty a crucial step to further progress.
Some advancements towards an accurate determination of the quarkonium wavefunctions at the origin also for strongly coupled quarkonia
has been made recently in refs.~\cite{Chung:2020zqc, Chung:2021efj}.
From a general perspective, it is important to emphasize that our phenomenological results, having been derived in an effective field theory framework, 
can be systematically improved through the formalism developed in this work by including corrections of higher order in $v$.
These come from higher dimensional operators in the NRQCD factorization formula,
from higher order corrections to the pNRQCD expansion of the NRQCD long-distance matrix elements,
and from higher order corrections to the wavefunctions originating from higher order corrections to the pNRQCD potential.
Higher order corrections in $\alpha_{\text{s}}$ can be included within perturbative QCD.
Studies of relativistic corrections to the NRQCD factorization formula from higher dimensional operators in refs.~\cite{He:2014sga, He:2015gla}
suggest that corrections of higher orders in $v$ can be appreciable for inclusive production of $J/\psi$,
although phenomenological applications have been limited so far by the lack of determinations of these higher order long-distance matrix elements.

As a first and obvious outlook, we expect the formalism developed in this work for computing production LDMEs to be applicable to any strongly coupled quarkonia,
not only the $P$-wave quarkonia considered here and in~\cite{Brambilla:2020ojz}.
These include the important case of $S$-wave quarkonia.
Calculations of the $S$-wave quarkonium production LDMEs may shed light on the longstanding puzzle of the $J/\psi$ production mechanism
(as long as the charmonium ground state may be assimilated to strongly coupled quarkonia),
as well as on issues in the polarization of $J/\psi$ and $\Upsilon(nS)$, and the $\eta_c$ production rate.
Further, it would be interesting to extend the formalism to describe the production of weakly coupled quarkonia,
which are possibly the lowest-lying quarkonium states, and also the production of quarkonium exotica (hybrids, tetraquarks).
In these cases, the formalism will require the introduction of nonperturbative matrix elements or functions that will be different from the ones encountered in the strong coupling regime.  
Finally, we also expect this approach to be useful in studying heavy quarkonium production in heavy ion collisions,
and help in this way to unveil the nature of the hot and dense phase of QCD.

\acknowledgments
We thank Geoffrey Bodwin for helpful comments.
The work of N.~B. is supported by the DFG (Deutsche Forschungsgemeinschaft, German Research Foundation) Grant No. BR 4058/2-2.
N.~B., H.~S.~C. and A.~V. acknowledge support from the DFG cluster of excellence ``ORIGINS'' under Germany's Excellence Strategy - EXC-2094 - 390783311.  
The work of A.~V. is funded by the DFG Project-ID 196253076 - TRR 110.

\bibliography{hadrolong.bib}
\bibliographystyle{JHEP}

\end{document}